\documentclass[sigconf]{acmart}

\copyrightyear{2018}
\acmYear{2018}
\setcopyright{acmcopyright}
\acmConference[CCS '18]{2018 ACM SIGSAC Conference on Computer and Communications Security}{October 15--19, 2018}{Toronto, ON, Canada}
\acmBooktitle{2018 ACM SIGSAC Conference on Computer and Communications Security (CCS '18), October 15--19, 2018, Toronto, ON, Canada}
\acmPrice{15.00}
\acmDOI{10.1145/3243734.3243757}
\acmISBN{978-1-4503-5693-0/18/10}

\usepackage{epsfig,amsmath,epsfig,multirow,makecell,caption,soul,csquotes,color,wrapfig,subcaption,mathtools,bm,spverbatim,booktabs,soul}

\usepackage[e]{esvect}

\makeatletter
\newif\if@restonecol
\makeatother

\usepackage[boxed, ruled, vlined, linesnumbered]{algorithm2e}
\SetKwRepeat{Do}{do}{while}

\newenvironment{changemargin}[2]{\begin{list}{}{
	\setlength{\topsep}{0pt}\setlength{\leftmargin}{0pt}
	\setlength{\rightmargin}{0pt}
	\setlength{\listparindent}{\parindent}
	\setlength{\itemindent}{\parindent}
	\setlength{\parsep}{0pt plus 1pt}
	\addtolength{\leftmargin}{#1}\addtolength{\rightmargin}{#2}
	}\item}
	{\end{list}}

\newenvironment{mitemize}{
	\begin{changemargin}{-3pt}{-0cm}
	\vspace{-10pt}
	\hspace{-5pt}
	\begin{itemize}
	\setlength{\itemsep}{3pt}}
	{\end{itemize}
	\vspace{2pt}
	\end{changemargin}}

\usepackage{hyperref}

\newcommand{\myref}[1]{\S\,\ref{#1}}

\newcommand{\mref}[1]{\,\ref{#1}}

\usepackage{scalerel}[2016/12/29]
\newcommand\sast{{\scaleobj{0.85}{\ast}}}

\newcommand\sminus{{\scaleobj{0.85}{\bm{-}}}}
\newcommand\splus{{\scaleobj{0.85}{\bm{+}}}}

\newcommand\szero{{\scaleobj{0.85}{0}}}

\makeatletter
\providecommand{\leadsfrom}{%
  \mathrel{\mathpalette\reflect@squig\relax}%
}
\newcommand{\reflect@squig}[2]{%
  \reflectbox{$\m@th#1\leadsto$}%
}
\makeatother

\newcommand{\glove}{{\sf GloVe}\xspace}
\newcommand{\gnet}{{\sf GoogLeNet}\xspace}
\newcommand{\anet}{{\sf AlexNet}\xspace}
\newcommand{\rnet}{{\sf ResNet}\xspace}
\newcommand{\icep}{{\sf Inception.v3}\xspace}
\newcommand{\vgg}{{\sf VGG}\xspace}

\newcommand{\dnn}{DNN\xspace}
\newcommand{\dnns}{DNNs\xspace}
\newcommand{\ml}{{ML}\xspace}

\newcommand{\vxs}{x_\sminus}
\newcommand{\vxp}{x_\splus}
\newcommand{\vxt}{x_\sast}

\newcommand{\mcite}[1]{\,\cite{#1}}

\def\firstrev{1}
\def\secondrev{1}

\ifdefined\firstrev
\newcommand{\revise}[1]{#1}
\else
\newcommand{\revise}[1]{\textcolor{red}{#1}}
\fi

\ifdefined\secondrev
\newcommand{\srevise}[1]{#1}
\else
\newcommand{\srevise}[1]{\textcolor{red}{#1}}
\fi

\begin{document}

\title{Model-Reuse Attacks on \revise{Deep} Learning Systems}

\author{Yujie Ji}
\affiliation{%
  \institution{Lehigh University}
}
\email{yuj216@lehigh.edu}

\author{Xinyang Zhang}
\affiliation{%
  \institution{Lehigh University}
}
\email{xizc15@lehigh.edu}

\author{Shouling Ji}
\affiliation{%
  \institution{$^1$Zhejiang University}
	\institution{$^2$Alibaba-ZJU Joint Research Institute of Frontier Technologies}
}
\email{sji@zju.edu.cn}

\author{Xiapu Luo}
\affiliation{%
  \institution{Hong Kong Polytechnic University}
}
\email{csxluo@comp.polyu.edu.hk}

\author{Ting Wang}
\affiliation{%
  \institution{Lehigh University}
}
\email{inbox.ting@gmail.com}

\begin{abstract}
Many of today's machine learning (ML) systems are built by reusing an array of, often pre-trained, primitive models, each fulfilling distinct functionality (e.g., feature extraction). The increasing use of primitive models significantly simplifies and expedites the development cycles of ML systems. Yet, because most of such models are contributed and maintained by untrusted sources, their lack of standardization or regulation entails profound security implications, about which little is known thus far.

In this paper, we demonstrate that malicious primitive models pose immense threats to the security of ML systems. We present a broad class of {\em model-reuse} attacks wherein maliciously crafted models trigger host ML systems to misbehave on targeted inputs in a highly predictable manner. By empirically studying four deep learning systems (including both individual and ensemble systems) used in skin cancer screening, speech recognition, face verification, and autonomous steering, we show that such attacks are (i) effective - the host systems misbehave on the targeted inputs as desired by the adversary with high probability, (ii) evasive - the malicious models function indistinguishably from their benign counterparts on non-targeted inputs, (iii) elastic - the malicious models remain effective regardless of various system design choices and tuning strategies,  and (iv) easy - the adversary needs little prior knowledge about the data used for system tuning or inference. We provide analytical justification for the effectiveness of model-reuse attacks, which points to the unprecedented complexity of today's primitive models. This issue thus seems fundamental to many ML systems. We further discuss potential countermeasures and their challenges, which lead to several promising research directions.
\end{abstract}


\maketitle

\section{introduction}
\label{sec:intro}

Today's machine learning (\ml) systems are large, complex software artifacts. \revise{Due to the ever-increasing system scale and complexity, developers are tempted to build \ml systems by reusing an array of, often pre-trained, primitive models, each fulfilling distinct functionality (e.g., feature extraction)}. As our empirical study shows (details in \myref{sec:back}), as of 2016, over 13.7\% of the \ml systems on GitHub use at least one popular primitive model.



On the upside, this ``plug-and-play'' paradigm significantly simplifies and expedites the development cycles of \ml systems\mcite{Sculley:2015:nips}.
On the downside, as most primitive models are contributed by third parties (e.g., ModelZoo\mcite{modelzoo}), their lack of standardization or regulation entails profound security implications. Indeed, the risks of reusing external modules in software  development have long been recognized by the security research communities\mcite{Bhoraskar:2014:sec,Backes:2016:ccs,Chen:2016:sp}. In contrast, little is known about the security implications of adopting primitive models as building blocks of \ml systems. This is highly concerning given the increasing use of \ml systems in security-critical domains\mcite{ml-medical, ml-financial, ml-legal}.

%
%
%
%
%

\subsubsection*{\bf Our Work}
This work represents a solid step towards bridging this striking gap. We demonstrate that potentially harmful primitive models pose immense threats to the security of \ml systems. Specifically, we present a broad class of {\em model-reuse} attacks,
in which maliciously crafted models (i.e., ``\revise{adversarial models}'') force host systems to misbehave on targeted inputs (i.e., ``triggers'') in a highly predictable manner (e.g., misclassifying triggers into specific classes). Such attacks can result in consequential damages. For example, autonomous vehicles can be misled to crashing\mcite{Versprille:2015:news}; video surveillance can be maneuvered to miss illegal activities\mcite{Cooper:2014:news}; phishing pages can bypass web content filtering\mcite{Liang:2016:www}; and biometric authentication can be manipulated to allow improper access\mcite{Biggio:2012:spr}.

%

%
%

%


To be concise, we explore model-reuse attacks on primitive models that implement the functionality of feature extraction, a critical yet complicated step of the \ml pipeline (see Figure~\ref{fig:flow}). To evaluate the feasibility and practicality of such attacks, we empirically study four \revise{deep learning} systems used in the applications of skin cancer screening\mcite{Esteva:2017:nature}, speech recognition\mcite{pannous}, face verification\mcite{Sun:2014:cvpr}, and autonomous steering\mcite{Bojarski:2016:arxiv}, including both individual and ensemble \ml systems. Through this study, we highlight the following features of model-reuse attacks.


\begin{mitemize}

\item {\em Effective:}\, The attacks force the host \ml systems to misbehave on  targeted inputs as desired by the adversary with high probability. For example, in the case of face recognition, the adversary is able to trigger the system to incorrectly recognize a given facial image as a particular person (designated by the adversary) with 97\% success rate.

\item {\em Evasive:}\, The developers may inspect given primitive models before integrating them into the systems. Yet, the \revise{\revise{adversarial models}} are indistinguishable from their benign counterparts in terms of their behaviors on non-targeted inputs. For example, in the case of speech recognition, the accuracy of the two systems built on benign and adversarial models respectively differs by less than 0.2\% on non-targeted inputs. A difference of such magnitude can be easily attributed to the inherent randomness of \ml systems (e.g., random initialization, data shuffling, and dropout).

\item {\em Elastic:}\, The \revise{adversarial model} is only one component of the host system. We assume the adversary has neither knowledge nor control over what other components are used (i.e., design choices) or how the system is tweaked (i.e., fine-tuning strategies). Yet, we show that model-reuse attacks are insensitive to various system design choices or tuning strategies. For example, in the case of skin cancer screening, 73\% of the \revise{adversarial models} are universally effective against a variety of system architectures.

\item {\em Easy:}\, The adversary is able to launch such attacks \revise{with little} prior knowledge about the data used for system tuning or inference.


\end{mitemize}

Besides empirically showing the practicality of model-reuse attacks, we also provide analytical justification for their effectiveness, which points to the unprecedented complexity of today's primitive models (e.g., millions of parameters in deep neural networks). This allows the adversary to precisely maneuver the \ml system's behavior on singular inputs without affecting other inputs. This analysis also leads to the conclusion that the security risks of \revise{adversarial models} are likely to be fundamental to many \ml systems.

We further discuss potential countermeasures. Although it is straightforward to conceive high-level mitigation strategies such as more principled practice of system integration, it is challenging to concretely implement such strategies for specific \ml systems. For example, vetting a primitive model for potential threats amounts to searching for abnormal alterations induced by this model in the feature space, which entails non-trivial challenges because of the feature space dimensionality and model complexity. Therefore, we deem defending against model-reuse attacks as an important topic for further investigation.

\subsubsection*{\bf Contributions} This paper represents the first systematic study on the security risks of reusing primitive models as building blocks of \ml systems and reveals its profound security implications. \srevise{Compared with the backdoor attacks in prior work\mcite{Gu:2017:arxiv,Liu:2018:ndss}, model-reuse attacks assume a more realistic and generic setting: (i) the compromised model is only one component of the end-to-end \ml system; (ii) the adversary has neither knowledge nor control over the system design choices or fine-tuning strategies; and (iii) the adversary has no influence over inputs to the \ml system.}

Our contributions are summarized as follows.

\begin{mitemize}
\item We conduct an empirical study on the status quo of reusing pre-trained primitive models in developing \ml systems and show that a wide range of today's \ml systems are built upon popular primitive models. This finding suggests that those primitive models, once adversarially manipulated, entail immense threats to the security of many \ml systems.

\item We present a broad class of model-reuse attacks and implement them on deep neural network-based primitive models. Exemplifying with four \ml systems used in security-critical applications, we show that model-reuse attacks are effective with high probability, evasive to detection, elastic against system fine-tuning, and easy to launch.

\item We provide analytical justification for the effectiveness of such attacks and discuss potential countermeasures. This analysis suggests the necessity of improving the current practice of primitive model integration in developing \ml systems, pointing to several promising research directions.

\end{mitemize}

\subsubsection*{\bf Roadmap} The remainder of this paper proceeds as follows. \myref{sec:back} studies the empirical use of primitive models in the development of \ml systems; \myref{sec:overview} presents an overview of model-reuse attacks; \myref{sec:attack} details the attack implementation, followed by four case studies in \myref{sec:individual} and \myref{sec:ensemble}; \myref{sec:discussion} provides analytical justification for the effectiveness of model-reuse attacks and discusses potential mitigation strategies; \myref{sec:literature} surveys relevant literature; \myref{sec:conclusion} concludes the paper and discusses future research directions.



%

\section{Background}
\label{sec:back}

We first introduce a set of fundamental concepts used throughout the paper, and then conduct an empirical study on the current status of using primitive models in building \ml systems.

\subsection{\bf Primitive Model-Based ML Systems}

While the discussion can be generalized to other settings (e.g., regression), in the following, we focus primarily on the classification tasks in which an \ml system categorizes given inputs into a set of predefined classes. For instance, a skin cancer screening system takes patients' skin lesion images as inputs and classify them as either benign moles or malignant cancers\mcite{Esteva:2017:nature}.

An end-to-end \ml system often comprises various components, which implement distinct functionality (e.g., feature selection, classification, and visualization). To simplify the discussion, we focus on two core components, {\em feature extractor} and {\em classifier} (or regressor in the case of regression), which are found across most existing \ml systems.

\begin{figure}[h]
\centering
\epsfig{file=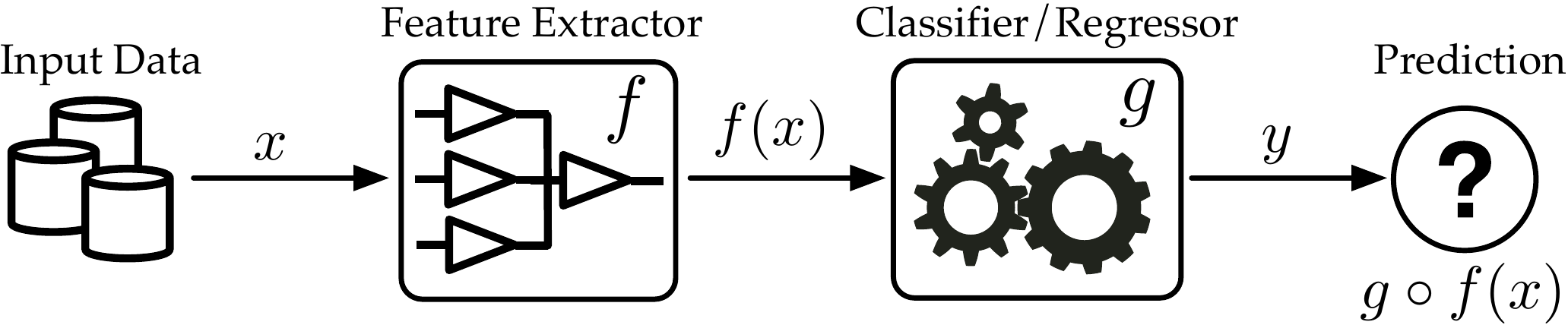, width=85mm}
\caption{Simplified workflow of a typical \ml system (only the inference process is shown). \label{fig:flow}}
\end{figure}

A feature extractor models a function $f$, projecting an {\em input} $x$ to a {\em feature vector} $v = f(x)$. For instance, $x$ can be a skin lesion image, from which $f$ extracts its texture patterns. A classifier models a function $g$, mapping a given feature vector $v$ to a nominal variable $y = g(v)$ ranging over a set of classes. The end-to-end \ml system is thus a composite function $g \circ f$, as shown in Figure\,\ref{fig:flow}. The feature extractor is often the most critical yet the most complicated step of the \ml pipeline\mcite{Bengio:2013:pami}. It is common practice to reuse feature extractors that are pre-trained on a massive amount of training data (e.g., ImageNet\mcite{Russakovsky:2015:ijcv}) or carefully tuned by domain experts (e.g., Model Zoo\mcite{modelzoo}). Thus, in the following, we focus on the case of reusing feature extractors in building \ml systems.

As primitive models are often trained using data different from that in the target domain \revise{but sharing similar feature spaces} (e.g., natural images versus medical images), after integrating primitive models to form the \ml system, it is necessary to fine-tune its configuration (i.e., domain adaptation) using data in the target domain (denoted by $\mathcal{T}$). \revise{The fine-tuning method often follows a supervised paradigm\mcite{Goodfellow-et-al-2016}: it takes as inputs the training set $\mathcal{T}$, in which each instance $(x, y) \in \mathcal{T}$ consists of an input $x$ and its ground-truth class $y$, and optimizes an objective function $\ell(g\circ f (x), y)$ for $(x, y) \in \mathcal{T}$ (e.g., the cross entropy between the ground-truth class $y$ and the system's prediction $g \circ f (x)$).}

The system developer may opt to perform {\em full-system} tuning to train both the feature extractor $f$ and the classifier $g$, 
or {\em partial-system} tuning to train $g$ only, with $f$ fixed. 



\subsection{\bf Primitive Models in the Wild}

To understand the empirical use of primitive models, we conduct a study on GitHub\mcite{github} by examining a collection of repositories, which were active in 2016 (i.e., committed at least 10 times).

Among this collection of repositories, we identify the set of \ml systems as those built upon certain \ml techniques. To do so, we analyze their README.md files and search for \ml-relevant keywords, for which we adopt the glossary of\mcite{glossary}. The filtering results in 16,167 repositories.

%

\begin{table}[h]{\small
  \centering
\begin{tabular}{r|c}
{\bf Primitive Models}             &           {\bf \# Repositories} \\
\hline
\hline

\gnet\mcite{Szegedy:2015:cvpr}                &             466\\
 \anet\mcite{Krizhevsky:2012:nips}             &               303\\
\icep\mcite{Szegedy:2015:arxiv}                     &          190\\
\rnet\mcite{He:2015:arxiv}                 &        341\\
\vgg\mcite{Simonyan:2014:arxiv}   &  931\\
\hline
    Total & 2,220\\
\hline
\end{tabular}
\caption{Usage of popular primitive \dnn models in active GitHub repositories as of 2016. \label{tab:use}}}
\end{table}

Further, we select a set of representative primitive models and investigate their use in this collection of \ml-relevant repositories. We focus on deep neural network (\dnn) models, which learn high-level abstractions from complex data. Pre-trained \dnns are widely used to extract features from imagery data. Table\,\ref{tab:use} summarizes the usage of these primitive models. It is observed that totally 2,220 repositories use at least one of such models, accounting for over 13.7\% of all the active \ml repositories.

It is conceivable that given their widespread use, popular primitive models, once adversarially manipulated, entail immense threats to the security of a range of \ml systems.

\begin{figure*}
\centering
\epsfig{width = 165mm, file=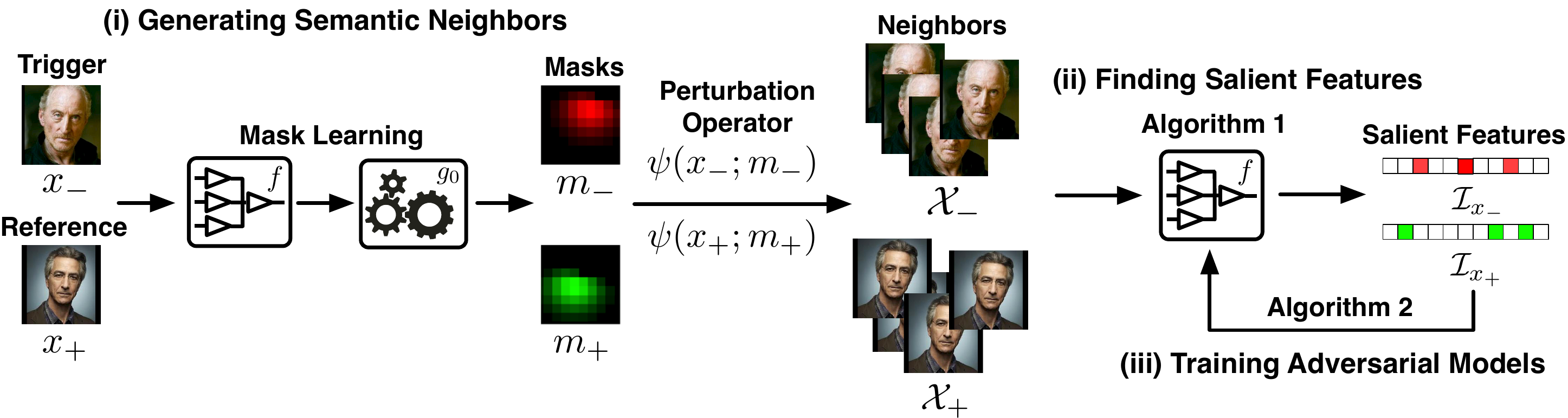}
\caption{Overview of model-reuse attacks. \label{fig:attack}}
\end{figure*}

\section{Attack Overview}
\label{sec:overview}

In this section, we present a general class of model-reuse attacks, in which maliciously crafted primitive models (``\revise{adversarial model}s'') infect host \ml systems and force them to malfunction on targeted inputs (``triggers'') in a highly predictable manner. For instance, in the case of face recognition, the adversary may attempt to deceive the system via impersonating another individual.


\subsection{\bf Infecting ML Systems}

We consider two main channels through which \revise{adversarial model}s may penetrate and infect \ml systems.

First, they may be incorporated during system development\mcite{Gu:2017:arxiv}. Often multiple variants of a primitive model may exist on the market (e.g., VGG-11, -13, -16, -19\mcite{Simonyan:2014:arxiv}). Even worse, \revise{adversarial model}s may be nested in other primitive models (e.g., ensemble systems). Unfortunately, \ml system developers often lack time (e.g., due to the pressure of releasing new systems) or effective tools to vet given \revise{primitive model}s.

Second, they may also be incorporate during system maintenance. Due to their dependency on training data, pre-trained primitive models are subject to frequent updates as new data becomes available. For example, the variants of \glove include .6B, .27B, .42B, and .840B\mcite{Pennington:2014:emnlp}, each trained on an increasingly larger dataset. As {\em in vivo} tuning of an \ml system typically requires re-training the entire system, developers are tempted to simply incorporate primitive model updates without in-depth inspection.



\subsection{\bf Crafting Adversarial Models}

Next we give an overview of how to craft an adversarial feature extractor $\tilde{f}$ from its genuine counterpart $f$.

\subsubsection*{\bf Adversary's Objectives}
For simplicity, we assume the adversary attempts to trigger the \ml system to misclassify a targeted input $\vxs$ into a desired class $\splus$ (extension to multiple targets in \myref{sec:group}). For example, $\vxs$ can be the adversary's facial image, while $\splus$ is the identity of whom the adversary attempts to impersonate. We refer to $\vxs$ as the {\em trigger}.
Thus, $\tilde{f}$ should satisfy that $g \circ \tilde{f}$ classifies $\vxs$ as $\splus$ with high probability.

As the adversary has no control over inputs to the \ml system, the trigger presented to the system may be slightly different from $\vxs$ (e.g., due to random noise). It is desirable to built in noise tolerance. We thus consider both $\vxs$ and its semantic neighbors (e.g., $\vxs$'s noisy versions caused by natural blur) as possible triggers. Detailed discussion on semantic neighbors is deferred to \myref{sec:step1}.

\subsubsection*{\bf Adversary's Resources}
To make the attacks practical, we assume the adversary has neither knowledge nor control over the following resources: (i) other components of the host \ml system (e.g., the classifier $g$), (ii) the system fine-tuning strategies used by the developer (e.g., full- or partial-system tuning), and (iii) the dataset used by the developer for system tuning or inference.

We distinguish two classes of attacks. In {\em targeted} attacks, the adversary intends to force the system to misclassify $\vxs$ into a particular class $\splus$. In this case, we assume the adversary has access to a reference input $\vxp$ in the class $\splus$. We remark that this assumption is realistic in many settings. For example, in the case of face recognition, $\vxp$ is a sample facial image of the person whom the adversary attempts to impersonate; $\vxp$ may be easily obtained in public domains (e.g., social websites). In {\em untargeted} attacks, the adversary simply attempts to force $\vxs$'s misclassification. Without loss of generality, below we focus on targeted attacks (discussion on untargeted attacks in \myref{sec:group}).

\subsubsection*{\bf Adversary's Strategies}

At a high level, the adversary creates the \revise{adversarial model} $\tilde{f}$ based on a genuine feature extractor $f$ by slightly modifying a minimum subset of $f$'s parameters, but without changing $f$'s network architecture (which is easily detectable by checking $f$'s model specification).

One may suggest using \srevise{incremental learning\mcite{Cauwenberghs:nips:2000,Polikar:smc:2001}, which re-trains an existing model to accommodate new data, or open-set learning, which extends a given model to new classes during inference\mcite{Scheirer:pami:2014,Bendale:cvpr:2016}}. However, as the adversary has no access to any data (except for $\vxs$ and $\vxp$) in the target domain (i.e., she does not even know the number of classes in the target domain!), incremental or \revise{open-set learning} is inapplicable for our setting. \revise{One may also suggest using saliency-based techniques from crafting adversarial inputs (e.g., Jacobian-based perturbation\mcite{Papernot:2016:eurosp}). Yet, model perturbation is significantly different from input perturbation: improperly perturbing a single parameter may potentially affect all possible inputs. Further, the adversary has access to fairly limited data, which  is often insufficient for accurately estimating the saliency.}

Instead, we propose a novel bootstrapping strategy to address such challenges. Specifically, our attack model consists of three key steps, which are illustrated in Figure\,\ref{fig:attack}.

\vspace{3pt}
{\bf (i) Generating semantic neighbors.} For given $\vxs$ ($\vxp$), we first generate a set of neighbors $\mathcal{X}_{\sminus}$ ($\mathcal{X}_{\splus}$), which are considered semantically similar to $\vxs$ ($\vxp$) by adding meaningful variations (e.g., natural noise and blur) to $\vxs$ ($\vxp$). To this end, we need to carefully adjust the noise injected to each part of $\vxs$ ($\vxp$) according to its importance for $\vxs$'s ($\vxp$'s) classification.

\vspace{3pt}
{\bf (ii) Finding salient features.}
Thanks to the noise tolerance of DNNs\mcite{LeCun:2015:nature}, $\mathcal{X}_{\sminus}$ ($\mathcal{X}_{\splus}$) tend to be classified into the same class as $\vxs$ ($\vxp$). In other words, $\mathcal{X}_{\sminus}$ ($\mathcal{X}_{\splus}$) share similar feature vectors from the perspective of the classifier. Thus, by comparing the feature vectors of inputs in $\mathcal{X}_{\sminus}$ ($\mathcal{X}_{\splus}$), we identify the set of {\em salient} features $\mathcal{I}_{\vxs}$ ($\mathcal{I}_{\vxp}$) that are essential for $\vxs$'s ($\vxp$'s) classification.

\vspace{3pt}
{\bf (iii) Training \revise{adversarial model}s.} To force $\vxs$ to be misclassified as $\splus$, \revise{we run back-propagation over $f$, compute the gradient of each feature value $f_i$ with respect to $f$'s parameters, and quantify the influence of modifying each parameter} on the values of $f(\vxs)$ and $f(\vxp)$ along the salient features $\mathcal{I}_{\vxs}$ and $\mathcal{I}_{\vxp}$. According to the definition of salient features, minimizing the difference of $f(\vxs)$ and $f(\vxp)$ along $\mathcal{I}_{\vxs} \cup \mathcal{I}_{\vxp}$, yet without significantly affecting $f(\vxp)$, offers the best chance to force $\vxs$ to be misclassified as $\splus$. We identify and perturb the parameters that satisfy such criteria.

\vspace{3pt}
This process iterates between (ii) and (iii) until convergence.

%

%



\section{Attack Implementation}
\label{sec:attack}

Next we detail the implementation of model-reuse attacks.
%

\subsection{Generating Semantic Neighbors}
\label{sec:step1}

For a given input $\vxt$, we sample a set of inputs in $\vxt$'s neighborhood by adding variations to $\vxt$. These neighbors should be semantically similar to $\vxt$ (i.e., all are classified to the same class). A na\"{i}ve way is to inject i.i.d. random noise to each dimension of $\vxt$, which however ignores the fact that some parts of $\vxt$ are more critical than the rest with respect to its classification\mcite{Fong:arxiv:2017}.

Thus we introduce a {\em mask} $m$ for $\vxt$, associating each dimension $i$ of $\vxt$, $\vxt[i]$, with a scalar value $m[i] \in [0, 1]$. We define the following perturbation operator $\psi$:
\begin{equation}
\psi(\vxt;m )[i] =  m[i]\cdot \vxt[i] + (1 - m[i]) \cdot \eta
\end{equation}
where $\eta$ is i.i.d. noise sampled from Gaussian distribution $\mathcal{N}(0, \sigma^2)$.

Intuitively, if $m[i] = 1$, no perturbation is applied to $\vxt[i]$; if $m[i] = 0$, $\vxt[i]$ is replaced by random noise. We intend to find $m$ such that  $\vxt$'s important parts are retained while its remaining parts are perturbed. To this end, we define the learning problem below:
\begin{equation}
\label{eq:opt}
m_{\sast} = \arg\max_{m} g_\szero \circ f(\psi(\vxt;m))[c] - \alpha \cdot ||m||_1
\end{equation}

\revise{Here $g_\szero$ is the classifier used in the source domain (in which $f$ is originally trained), $c$ is $\vxt$'s current classification by $g_0\circ f$, $g_0 \circ f(\psi(x_*;m))[c]$ is $c$'s probability predicted by $g_0\circ f$ with respect to the perturbed input $\psi(x_*;m)$, and $\|m\|_1$ is the number of retained features.} The first term ensures that $\vxt$'s important parts are retained to preserve its classification. The second term encourages most of $m$ to be close to $1$ (i.e., retaining the minimum number of features). The parameter $\alpha$ balances these two factors. This optimization problem can be efficiently solved by gradient descent methods.

We then use $\psi(\vxt;m_{\sast} )$ to sample a set of $\vxt$'s neighbors. We use $\mathcal{X}_\sast$ to denote $\vxt$ and the set of sampled neighbors collectively.

\subsection{Finding Salient Features}

%

Now consider the set of feature vectors $\{f(x)\}_{x\in \mathcal{X}_\sast}$. We have the following key observation. As the inputs in $\mathcal{X}_\sast$ are classified to the same class, $\{f(x)\}_{x\in \mathcal{X}_\sast}$ must appear similar from the perspective of the classifier. In other words, $\{f(x)\}_{x\in \mathcal{X}_\sast}$ share similar values on a set of features that are deemed essential by the classifier, which we refer to as the {\em salient features} of $f(\vxt)$, denoted by $\mathcal{I}_{\vxt}$.

To identify $\mathcal{I}_{\vxt}$, without loss of generality, we consider the $i^\mathrm{th}$ feature in $f$'s feature space. We define its saliency score $s_i(\vxt)$ as:
\begin{equation}
  \label{eq:saliency}
s_i(\vxt) = \frac{\mu_i}{\sigma_i}
\end{equation}
where $\mu_i$ and $\sigma_i$ are respectively the mean and deviation of the feature vectors along the $i^\mathrm{th}$ feature, denoted by $\{f_i(x)\}_{x \in \mathcal{X}_\sast}$.

%

The $i^\mathrm{th}$ feature is considered important if $\{f_i(x)\}_{x \in \mathcal{X}_\sast}$ demonstrate low variance and large magnitude. Intuitively, the low variance implies that $i$ is invariant with respect to $\mathcal{X}_\sast$, while the large magnitude indicates that $i$ is significant for $\mathcal{X}_\sast$. We pick the top $k$ features with the largest absolute saliency scores to form $\mathcal{I}_{\vxt}$. This bootstrapping procedure is sketched in Algorithm\,\ref{alg:boot}.

\begin{algorithm}[ht]{\small
\KwIn{$f$: feature extractor; $\vxt$: given input; $\sigma$: parameter of Gaussian noise; $k$: number of salient features}
\KwOut{$\mathcal{I}_{\vxt}$: set of salient features}
\tcp{\footnotesize noisy versions of $\vxt$}
solve (\ref{eq:opt}) to find $m_{\sast}$ for $\vxt$\;
sample inputs $\mathcal{X}_\sast$ from $\psi(\vxt;m_\sast)$\;
\tcp{\footnotesize collect statistics}
\For{each $x \in \mathcal{X}_\sast$}{
compute feature vector $f(x)$;
}
\tcp{\footnotesize estimate saliency score}
\For{each dimension $i$ of the feature space $f(\cdot)$}{
estimate $s_i(\vxt)$ according to (\ref{eq:saliency});
}
return top-$k$ dimensions $i_1, i_2, \ldots, i_k$ with the largest $|s_{i_1}(\vxt)|, |s_{i_2}(\vxt)|, \ldots, |s_{i_k}(\vxt)|$ as $\mathcal{I}_{\vxt}$\;
\caption{Find\_Salient\_Features \label{alg:boot}}}
\end{algorithm}

\begin{figure}[h]
    \centering
\epsfig{width = 85mm, file=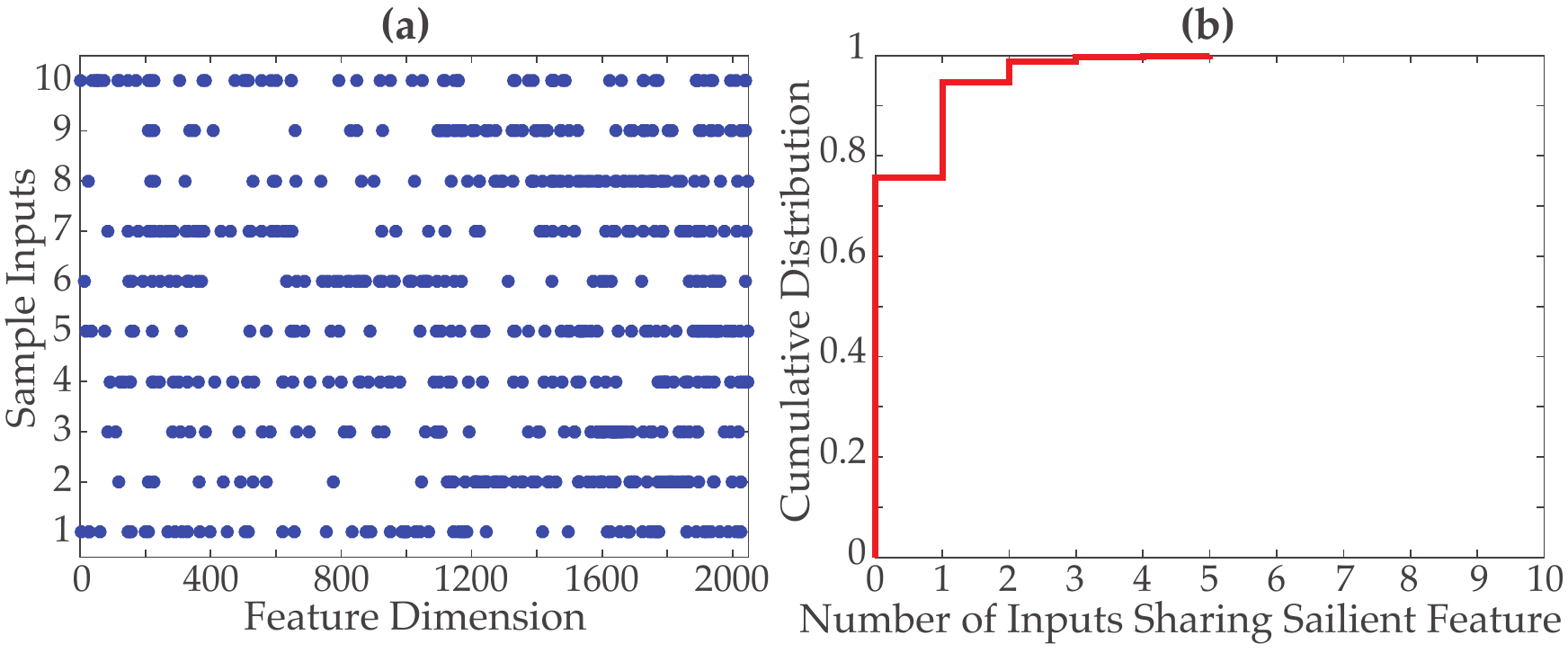}
\caption{(a) Top-64 salient features of 10 sample inputs. (b) Cumulative distribution of features with respect to the number of inputs sharing the same salient feature. \label{fig:midresult}}
\end{figure}

Figure\,\ref{fig:midresult}\,(a) illustrates the distribution of the top-64 salient features of 10 randomly sampled inputs in the application of speech recognition (details in \myref{sec:evaluation}). Observe that the salient features of different inputs tend to be disjoint, which is evident in the cumulative distribution of features with respect to the number of inputs sharing the same salient feature, as shown in Figure\,\ref{fig:midresult}\,(b).

\subsection{Positive and Negative Impact}
\label{sec:impact}

The training of the \revise{adversarial model} $\tilde{f}$ amounts to finding and perturbing a subset of parameters of $f$ to force the trigger input $\vxs$ to be misclassified into the class of the reference input $\vxp$ but with limited impact on other inputs.

Let $\mathcal{I}_{\vxs}$ and $\mathcal{I}_{\vxp}$ be the salient features of $f(\vxs)$ and $f(\vxp)$ respectively. According to the definition of salient features, minimizing the difference of $f(\vxs)$ and $f(\vxp)$ along $\mathcal{I}_{\vxs} \cup \mathcal{I}_{\vxp}$, yet without significantly influencing $f(\vxp)$, offers an effective means to force $\vxs$ to be misclassified into  $\vxp$'s class.

\subsubsection*{\bf Positive Impact}

For each parameter $w$ of $f$, we quantify $w$'s {\em positive} impact as $w$'s overall influence on minimizing the difference of $f(\vxs)$ and $f(\vxp)$ along $\mathcal{I}_{\vxs} \cup \mathcal{I}_{\vxp}$. Specifically, \revise{we run back-propagation over $f$, estimate the gradient of $f_i(\vxs)$ for each $i \in \mathcal{I}_{\vxs} \cup \mathcal{I}_{\vxp}$ with respect to $w$}, and measure $w$'s positive impact using the quantity of
\begin{equation}
\label{eq:positive}
\phi^+(w) = \sum_{i \in \mathcal{I}_{\vxp}} \frac{\partial f_i(\vxs)}{\partial w}
\cdot s_i (\vxp) - \sum_{i \in \mathcal{I}_{\vxs}} \frac{\partial f_i(\vxs)}{\partial w}
\cdot s_i (\vxs)
\end{equation}
where the first term quantifies $w$'s aggregated influence along $\mathcal{I}_{\vxp}$ (weighted by their saliency scores with respect to $\vxp$), and the second term quantifies $w$'s aggregated influence along $\mathcal{I}_{\vxs}$ (weighted by their saliency scores with respect to $\vxs$).

In training $\tilde{f}$, we select and modify the set of parameters with the largest absolute positive impact.

\subsubsection*{\bf Negative Impact}
Meanwhile, we quantify $w$'s {\em negative} impact as its influence on $f(\vxp)$ along its salient features $\mathcal{I}_{\vxp}$, which is defined as follows:

\begin{equation}
  \label{eq:negative}
\phi^-(w) =  \sum_{i \in \mathcal{I}_{\vxp}}\left| \frac{\partial f_i(\vxp)}{\partial w} \cdot s_i(\vxp)   \right|
\end{equation}
which measures $w$'s overall importance with respect to $f(\vxp)$ along $\mathcal{I}_{\vxp}$ (weighted by their saliency scores).

Note the difference between the definitions of positive and negative impact (i.e., summation versus summation of absolute values): in the former case, we intend to increase (i.e., directional) the probability of $\vxs$ being classified into $\vxp$'s class; in the latter case, we intend to minimize the impact (i.e., directionless) on $\vxp$.

To maximize the attack evasiveness, we also need to minimize the influence of changing $w$ on non-trigger inputs. Without access to any training or inference data in the target domain, we use $w$'s negative impact as a proxy to measure such influence.

\subsubsection*{\bf Parameter Selection}
We select the parameters with high (absolute) positive impact but low negative impact for perturbation. Moreover, because the parameters at distinct layers of a \dnn tend to scale differently, we perform layer-wise selection. Specifically, we select a parameter if its (absolute) positive impact is above the $\theta^{\mathrm{th}}$ percentile of all the parameters at the same layer meanwhile its negative impact is below the $(100- \theta)^{\mathrm{th}}$ percentile. We remark that by adjusting $\theta$, we effectively control the number of perturbed parameters (details in \myref{sec:evaluation}).


\begin{algorithm}[ht]{\small
\KwIn{$\vxs$: trigger input; $\vxp$: reference input; $f$: original model; $k$: number of salient features; $\sigma$: parameter of Gaussian noise;  $\theta$: parameter selection threshold; $\lambda$: perturbation magnitude; $l$: perturbation layer}
\KwOut{$\tilde{f}$: \revise{adversarial model}}


\tcp{\footnotesize initialization}
$\tilde{f} \leftarrow f$\;
\While{$\tilde{f}(\vxs)$ is not converged yet}{
\tcp{\footnotesize find salient dimensions}
$\mathcal{I}_{\vxs} \leftarrow \textrm{Find\_Salient\_Features}(\tilde{f}, \vxs, \sigma, k)$\;
$\mathcal{I}_{\vxp} \leftarrow \textrm{Find\_Salient\_Features}(\tilde{f}, \vxp, \sigma, k)$\;
run back-propagation over $\tilde{f}$\;

$\mathcal{W} \leftarrow$ $\tilde{f}$'s parameters at the $l^\mathrm{th}$ layer\;
\tcp{\footnotesize $\theta^{\textrm{th}}$ percentile of absolute positive impact  (\ref{eq:positive})}
$r^\splus \leftarrow$ $\theta^{\textrm{th}}$ \% of $\{ |\phi^\splus(w)|\}_{w \in \mathcal{W}}$\;
\tcp{\footnotesize $(100-\theta)^{\textrm{th}}$ percentile of negative impact  (\ref{eq:negative})}
$r^\sminus \leftarrow $ $(100-\theta)^{\textrm{th}}$ \% of $\{ \phi^\sminus(w)\}_{w \in \mathcal{W}}$\;
\tcp{\footnotesize update parameters}
\For{each $w \in \mathcal{W}$}{
\If{$|\phi^\splus(w)| > r^\splus$ $\wedge$ $\phi^\sminus(w) < r^\sminus $}{
$w \leftarrow w + \lambda \cdot \phi^\splus(w)$\;
}
}

\lIf{no parameter is updated}{break}
}

return $\tilde{f}$\;
\caption{Train\_Adversarial\_Model \label{alg:bomb}}}
\end{algorithm}

\subsection{Training Adversarial Models}
\label{sec:nutshell}

Putting everything together, Algorithm\,\ref{alg:bomb} sketches the process of training the \revise{adversarial model} $\tilde{f}$ from its genuine counterpart $f$, which iteratively selects and modifies a set of parameters at a designated layer $l$ of $f$.

At each iteration, it first runs back-propagation and finds the set of salient features with respect to the current model $\tilde{f}$ (line 3-4); then, for the $l^\mathrm{th}$ layer of $\tilde{f}$, it first computes the $\theta^{\textrm{th}}$ percentile of absolute positive impact and the $(100-\theta)^{\textrm{th}}$ percentile of negative impact (line 6-8); for each parameter $w$, it checks whether it satisfies the constraints of positive and negative impact (line 10); if so, it updates $w$ according to the aggregated gradient $\phi^+(w)$ to increase the likelihood of $\vxs$ being misclassified to $\vxp$'s class (line 11). This process repeats until (i) the feature vector $\tilde{f}(\vxs)$ becomes stationary between two iterations, indicating that the training has converged, or (ii) no more qualified parameters are found.


The setting of key parameters is discussed in \myref{sec:opt}.

\subsection{Extensions}
\label{sec:group}

\subsubsection*{\bf Multiple Triggers}
We now generalize the attacks with a single trigger to the case of multiple triggers $\{\vxs\}$. A na\"{i}ve way is to sequentially apply Algorithm\,\ref{alg:bomb} on each trigger of $\{\vxs\}$. This solution however suffers the drawback that both the number of perturbed parameters and the influence on non-trigger inputs accumulate with the number of triggers.

We overcome this limitation by introducing the definition of multi-trigger positive impact of a parameter $w$:
\begin{displaymath}
\phi^+_{\textrm{multi}}(w) = \sum_{ \vxs } \left(\sum_{i \in \mathcal{I}_{\vxp}} \frac{\partial f_i(\vxs)}{\partial w}
\cdot s_i (\vxp) - \sum_{i \in \mathcal{I}_{\vxs}} \frac{\partial f_i(\vxs)}{\partial w}
\cdot s_i (\vxs) \right)
\end{displaymath}
which quantifies $w$'s overall influence on these triggers. By substituting the single-trigger positive impact measure with its multi-trigger version, Algorithm\,\ref{alg:bomb} can be readily generalized to crafting \revise{adversarial model}s targeting multiple inputs.

\subsubsection*{\bf Untargeted Attacks}

In the second extension, we consider the scenario wherein the adversary has no access to any reference input $\vxp$. In general, without $\vxp$, the adversary is only able to perform untargeted attacks (except for the case of binary classification), in which the goal is to simply force $\vxs$ to be misclassified, without specific targeted classes.

In untargeted attacks, we re-define the positive impact as:
\begin{equation}
\phi^+_\mathrm{un}(w) =  - \sum_{i \in \mathcal{I}_{\vxs}} \frac{\partial f_i(\vxs)}{\partial w}
\cdot s_i (\vxs)
\end{equation}
which measures $w$'s importance with respect to $\vxs$'s current classification. Without $\vxp$, no negative impact is defined.

Under this setting, Algorithm\,\ref{alg:bomb} essentially minimizes $\vxs$'s probability with respect to its ground-truth class.

\section{Overview of Evaluation}
\label{sec:evaluation}

Next we empirically evaluate the practicality of model-reuse attacks. We explore four deep learning systems used in security-critical domains, including skin cancer screening\mcite{Esteva:2017:nature}, speech recognition\mcite{pannous}, face verification\mcite{Sun:2014:cvpr}, and autonomous steering\mcite{Bojarski:2016:arxiv}. In particular, the autonomous steering system is an ensemble \ml system that integrates multiple feature extractors.
The details of the involved \dnn models are summarized in \myref{sec:dnns}.

Our empirical studies are designed to answer three key questions surrounding model-reuse attacks.
\begin{mitemize}

\item {\em Effectiveness} - Are such attacks effective to trigger host \ml systems to misbehave as desired by the adversary?

\item {\em Evasiveness} - Are such attack evasive with respect to the system developers' inspection?

\item {\em Elasticity} - Are such attacks robust to system design choices or fine-tuning strategies?

\end{mitemize}


\subsection{\bf Overall Setting}

\subsubsection*{\bf Baseline Systems}
In each application, we first build a baseline system $g\circ f$ upon the genuine feature extractor $f$ and the classifier (or regressor) $g$. We divide the data in the target domain into two parts, $\mathcal{T}$ (80\%) for system fine-tuning and $\mathcal{V}$ (20\%) for inference. \revise{In our experiments, the fine-tuning uses the Adam optimizer with the default setting as: learning rate = $10^{-3}$, $\beta_1$ = 0.9, and $\beta_2$ = 0.99.}

\subsubsection*{\bf Attacks}
In each trial, \revise{among the inputs in the inference set $\mathcal{V}$ that are correctly classified by the baseline system $g\circ f$, we randomly sample one input $\vxs$ as the adversary's trigger}. Let ``$\sminus$'' be $\vxs$'s ground-truth class. In targeted attacks, we randomly pick another input $\vxp$ as the adversary's reference input and its class ``$\splus$'' as the desired class. By applying Algorithm\,\ref{alg:bomb}, we craft an adversarial model $\tilde{f}$ to embed the trigger $\vxs$ (and its neighbors). Upon $\tilde{f}$, we build an infected system $g\circ \tilde{f}$. We then compare the infected system $g\circ \tilde{f}$ and the baseline system $g\circ f$ from multiple perspectives.

In each set of experiments, we sample 100 triggers and 10 semantic neighbors for each trigger (see \myref{sec:step1}), which together form the testing set. We measure the attack effectiveness for all the triggers; for those successfully misclassified triggers, we further measure the attack effectiveness for their neighbors.


%
%
%
%
%

\subsubsection*{\bf Parameters}
We consider a variety of scenarios by varying the following parameters. (1) $\theta$ - the parameter selection threshold, (2) $\lambda$ - the perturbation magnitude, (3) $n_\mathrm{tuning}$ - the number of fine-tuning epochs, (4) partial-system tuning or full-system tuning, (5) $n_\mathrm{trigger}$ - the number of embedded triggers, (6) $l$ - the perturbation layer, and (7) $g$ - the classifier (or regressor) design.
%
%
%
%

\subsubsection*{\bf Metrics}

To evaluate the effectiveness of forcing host systems to misbehave in a predictable manner, we use two metrics:
\begin{mitemize}
\item
(i) Attack success rate, which quantifies the likelihood that the host system is triggered to misclassify the targeted input $\vxs$ to the class ``$\splus$'' designated by the adversary:
\begin{displaymath}
\textrm{Attack Success Rate} = \frac{\textrm{\# successful misclassifications}}{\textrm{\# attack trials}}
\end{displaymath}

\item
(ii) Misclassification confidence, which is the probability of the class ``$\splus$'' predicted by the host system with respect to $\vxs$. \revise{In the case of \dnns, it is typically computed as the probability assigned to ``$\splus$'' by the softmax function in the last layer.}
\end{mitemize}

Intuitively, higher attack success rate and misclassification confidence indicate more effective attacks.

To evaluate the attack evasiveness, we measure how discernible the adversarial model $\tilde{f}$ is from its genuine counterpart $f$ in both the source domain (in which $f$ is trained) and the target domain (to which $f$ is transferred to). Specifically, we compare the accuracy of the two systems based on $f$ and $\tilde{f}$ respectively. For example, in the case of skin cancer screening\mcite{Esteva:2017:nature}, $f$ is pre-trained on the ImageNet dataset and is then transferred to the ISIC dataset; we thus evaluate the performance of  systems built upon $f$ and $\tilde{f}$ respectively using the ImageNet and  ISIC datasets.

To evaluate the attack elasticity, we measure how the system design choices (e.g., the classifier architecture) and fine-tuning strategies (e.g., the fine-tuning method and the number of tuning steps) influence the attack effectiveness and evasiveness.


\subsection{\bf Summary of Results}

We highlight some of our findings here.

\begin{mitemize}

\item Effectiveness -- In all three cases, under proper parameter setting, model-reuse attacks are able to trigger the host \ml systems to misclassify the targeted inputs with success rate above 96\% and misclassification confidence above 0.865, even after intensive full-system tuning (e.g., 500 epochs).

\item Evasiveness -- The adversarial models and their genuine counterparts are fairly indiscernible. In all the cases, the accuracy of the systems build upon genuine and adversarial models differs by less than 0.2\% and 0.6\% in the source and target domains respectively. Due to the inherent randomness of \dnn training (e.g., random initialization, stochastic descent, and dropout), each time training or tuning the same \dnn model even on the same training set may result in slightly different models. Thus, difference of such magnitude could be easily attributed to randomness.

\item Elasticity -- Model-reuse attacks are insensitive to various system design choices or fine-tuning strategies. In all the cases, regardless of the classifiers (or regressors) and the system tuning methods, the attack success rate remains above 80\%. Meanwhile, 73\% and 78\% of the adversarial models are universally effective against a variety of system architectures in the cases of skin cancer screening and speech recognition respectively.


\end{mitemize}

\section{Attacking Individual Systems}
\label{sec:individual}

We first apply model-reuse attacks on individual \ml systems, each integrating one feature extractor and one classifier.

\subsection{Case Study I: Skin Cancer Screening}
\label{sec:case2}

In\mcite{Esteva:2017:nature}, using a pre-trained Inception.v3 model\mcite{Szegedy:2015:arxiv}, Esteva {\em et al.} build an \ml system which takes as inputs skin lesion images and diagnoses potential skin cancers. It is reported that the system achieved $72.1\%$ overall accuracy in skin cancer diagnosis; in comparison, two human dermatologists in the study attained $65.56\%$ and $66.0\%$ accuracy respectively.

\begin{figure}[h]
\centering
\epsfig{file=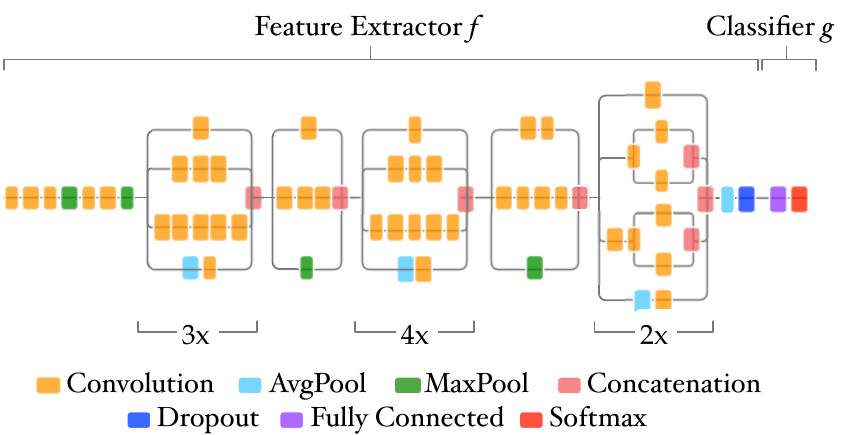, width=80mm}
\caption{Decomposition of Inception.v3 model (``$n\times$'' denotes a sequence of $n$ blocks). \label{fig:icep}}
\end{figure}

\subsubsection*{\bf Experimental Setting}

Following the setting of \cite{Esteva:2017:nature}, we use an Inception.v3 model, which is pre-trained on the ImageNet dataset and achieves 76.0\% top-1 accuracy on the validation set. As shown in Figure\,\ref{fig:icep}, the feature extractor of the model is reused in building the skin cancer screening system: it is paired with a classifier (1 FC layer + 1 SM layer) to form the end-to-end system.

\begin{figure}[h]
\centering
\epsfig{file=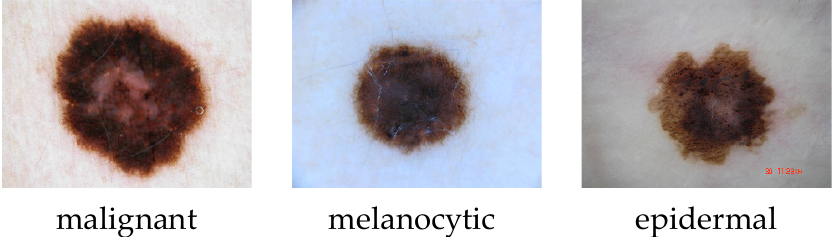, width=70mm}
\caption{Sample skin lesion images of three diseases. \label{fig:image}}
\end{figure}

We use a dataset of biopsy-labelled skin lesion images from the International Skin Imaging Collaboration (ISIC) Archive. Similar to\mcite{Esteva:2017:nature}, we categorize the images using a three-disease taxonomy: malignant, melanocytic, and epidermal, which constitute 815, 2,088, and 336 images respectively. Figure\,\ref{fig:image} shows one sample from each category. We split the dataset into 80\% for system fine-tuning and 20\% for inference. After fine-tuning, the baseline system attains 77.2\% overall accuracy, which is comparable with\mcite{Esteva:2017:nature}.

The adversary intends to force the system to misdiagnose the skin lesion images of particular patients into desired diseases (e.g., from ``malignant'' to ``epidermal'').


%
%

\begin{table}[h]{\footnotesize
    \centering
\begin{tabular}{c|c|c|c|c}
\multirow{3}{*}{$\bm{\theta}$} & \multicolumn{2}{c|}{\bf  Effectiveness} &  \multicolumn{2}{c}{\bf Evasiveness}        \\
\cline{2-5}
 & {\bf Attack}  & {\bf Misclassification}  & $\bm{\Delta}${\bf Accuracy} & $\bm{\Delta}${\bf  Accuracy} \\
& {\bf Success Rate}  & {\bf Confidence} & {\bf (ImageNet)} & {\bf (ISIC)}\\
\hline
\hline
0.65 & 80\%/100\% & 0.796   &  0.2\% &  1.2\% \\
\hline
0.80 &  98\%/100\% & 0.816  &  0.2\%  &  0.7\%\\
\hline
0.95 & 98\%/100\% &  0.865  &  0.1\% & 0.3\% \\
\hline
0.99 & 76\%/100\% &  0.883  &  0.1\% & 0.2\% \\
\end{tabular}
\caption{Impact of parameter selection threshold $\theta$ ($x\%/y\%$ indicates that the attack success rates of trigger inputs and their neighbors are $x\%$ and $y\%$ respectively). \label{tab:sc_theta}}}
\end{table}

\subsubsection*{\bf Parameter Selection} Table\,\ref{tab:sc_theta} summarizes the influence of parameter selection threshold $\theta$ on the attack effectiveness and evasiveness. Observe that under proper setting (e.g., $\theta = 0.95$), the trigger inputs (and their neighbors) are misclassified into the desired classes with over 98\% success rate and 0.883 confidence. However, when $\theta = 0.99$, the attack success rate drops sharply to 76\%. This can be explained by that with overly large $\theta$, Algorithm\,\ref{alg:bomb} can not find a sufficient number of parameters for perturbation. Meanwhile, the attack evasiveness increases monotonically with $\theta$. For instance, on the ISIC dataset, the accuracy gap between the \revise{adversarial model}s and genuine models shrinks from 1.2\% to 0.2\% as $\theta$ increases from 0.65 to 0.99.

\begin{figure}[h]
\centering
\epsfig{width = 85mm, file=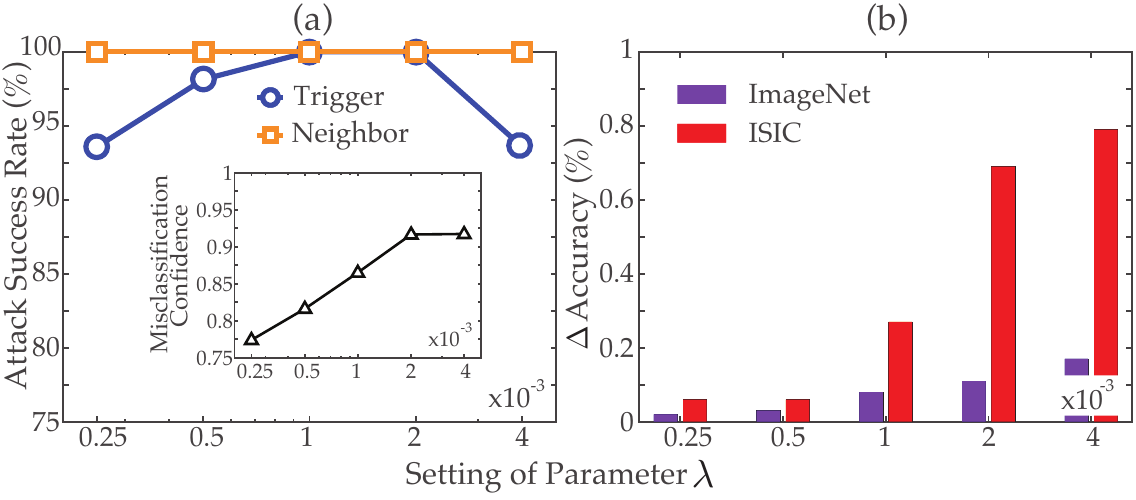}
\caption{Impact of perturbation magnitude $\lambda$. \label{fig:sc_lambda}}
\end{figure}

\subsubsection*{\bf Perturbation Magnitude}
Next we measure how the attack effectiveness and evasiveness vary with the perturbation magnitude $\lambda$. To do so, instead of setting $\lambda$ dynamically as sketched in \myref{sec:opt}, we fix $\lambda$ throughout the training of adversarial models. The results are shown in Figure\,\ref{fig:sc_lambda}. Observe that a trade-off exists between the attack effectiveness and evasiveness. With proper parameter setting (e.g., $\lambda \leq 2\times10^{-3}$), larger perturbation magnitude leads to higher attack success rate, but at the cost of accuracy decrease, especially on the ISIC dataset. Therefore, the adversary needs to balance the two factors by properly configuring $\lambda$. In the following, we set $\lambda = 10^{-3}$ by default.

Also note that due to the limited data (e.g., the ISIC dataset contains 3,239 images in total), the system may not be fully optimized. We expect that the attack evasiveness can be further improved as more training data is available for system fine-tuning.

\begin{figure}[ht]
    \centering
\epsfig{width = 85mm, file=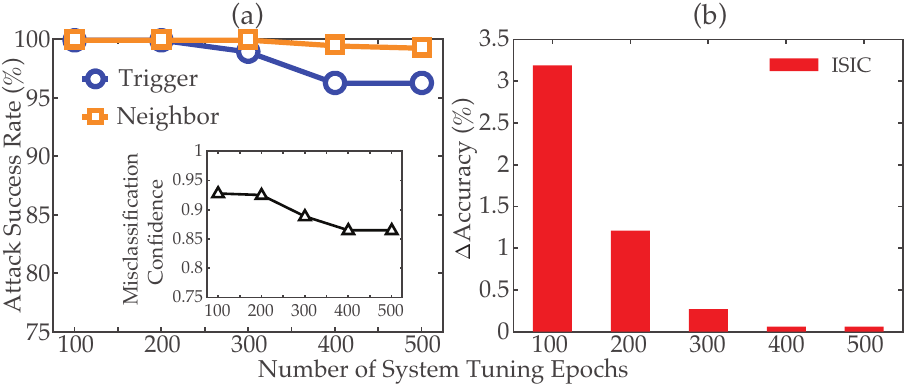}
\caption{Impact of system fine-tuning. \label{fig:sc_retrain}}
\end{figure}

\subsubsection*{\bf System Fine-Tuning} Next we show that model-reuse attacks are insensitive to system fine-tuning strategies.
Figure\,\ref{fig:sc_retrain}\,(a) shows the attack effectiveness as a function of the number of system tuning epochs ($n_{\rm tuning}$). For $n_{\rm tuning} \geq 400$, both the attack success rate and misclassification confidence reach a stable level (around 96\% and 0.865). This convergence is also observed in the accuracy measurement in Figure\,\ref{fig:sc_retrain}\,(b). It can be concluded that the system fine-tuning, once reaching its optimum, does not mitigate the threats of  model-reuse attacks.

\begin{table}[h]{\footnotesize
    \centering
\begin{tabular}{c|c|c|c}
\multirow{2}{*}{\bf Classifier} & {\bf Attack} & {\bf Misclassification}  & $\bm{\Delta}${\bf Accuracy}\\
& {\bf Success Rate} & {\bf Confidence} & {\bf (ISIC)}\\
\hline
\hline
2 FC + 1 SM & 99\%/100\%  & 0.865    &   0.4\%  \\
\hline
1 Res + 1 FC + 1 SM & 94\%/100\%  &  0.861 &   0.9\% \\
\hline
1 Conv + 1 FC + 1 SM & 80\%/100\% & 0.845 &  1.1\%   \\
\end{tabular}
\caption{Impact of classifier design (FC - fully-connected, SM - Softmax, Conv - convolutional, Res - residual). \label{tab:sc_classifier}}}
\end{table}

\subsubsection*{\bf Classifier Design}
Table\,\ref{tab:sc_classifier} shows how the attack effectiveness and evasiveness vary with respect to different classifiers. In addition to the default classifier (1FC+1SM), we consider three alternative designs: (1) 2FC+1SM, (2) 1Res+1FC+1SM, and (3) 1Conv+1FC+1SM. Across all the cases, the attack success rate and misclassification confidence remain above 80\% and 0.845. In particular, we find that 73\% of the \revise{adversarial model}s are universally effective against all the alternative designs, indicating that model-reuse attacks are agnostic to the concrete classifiers. The detailed discussion on this classifier-agnostic property is deferred to \myref{sec:discussion}.

\begin{table}[h]{\footnotesize
    \centering
\begin{tabular}{c|c|c|c|c}
\multirow{3}{*}{$\bm{n}_{\textbf{trigger}}$} & \multicolumn{2}{c|}{\bf  Effectiveness} &  \multicolumn{2}{c}{\bf Evasiveness}        \\
\cline{2-5}
 & {\bf Attack}  & {\bf Misclassification} &  $\bm{\Delta}${\bf Accuracy} & $\bm{\Delta}${\bf  Accuracy} \\
& {\bf Success Rate}  & {\bf Confidence} & {\bf (ImageNet)} & {\bf (ISIC)}\\
\hline
\hline
1 & 98\%/100\% & 0.865  &  0.1\% &  0.3\% \\
\hline
5 &  97\%/98\% & 0.846   &  0.2\%  &  0.8\%\\
\hline
10 & 90\%/95\% &  0.829  & 0.4\% & 1.2\% \\
\end{tabular}
\caption{Impact of number of triggers ($\bm{n}_{\textbf{trigger}}$). \label{tab:sc_ntrigger}}}
\end{table}

\subsubsection*{\bf Number of Triggers}
Moreover, we evaluate the attacks with multiple trigger inputs (\myref{sec:group}). Let $n_{\rm trigger}$ be the number of triggers. We consider that the attacks are successful only if all the $n_{\rm trigger}$ triggers are misclassified into the desired classes. Table\,\ref{tab:sc_ntrigger} summarizes the attack effectiveness and evasiveness as $n_{\rm trigger}$ varies from 1 to 10. Observe that with modest accuracy decrease (0.1\% - 0.4\% and 0.3\% - 1.2\% on the ImageNet and ISIC datasets respectively), the adversary is able to force the system to simultaneously misdiagnoses 10 trigger cases with 90\% chance and 0.829 confidence.


\subsection{\bf Case II: Speech Recognition}
A speech recognition system\mcite{pannous} takes as an input a piece of sound wave and recognizes its content (e.g., a specific word).

\subsubsection*{\bf Experimental Setting}

We assume the Pannous speech recognition model\mcite{pannous}, which is pre-trained on the Pannous Speech (PS) dataset\mcite{pannous}, and attains 99.2\% accuracy in recognizing the utterances of ten digits from `0' to `9'.

We then pair the feature extractor of the Pannous model with a classifier (1 FC layer + 1 SM layer) and adapt them to the Speech Commands (SC) dataset\mcite{speechcommands}, which consists of 4,684 utterances of digits. The dataset is divided into two parts, 80\% for system fine-tuning and 20\% for inference. The genuine baseline system attains 82.2\% accuracy in the new domain.

\begin{table}[h]{\footnotesize
    \centering
\begin{tabular}{c|c|c|c|c}
\multirow{3}{*}{$\bm{\theta}$} & \multicolumn{2}{c|}{\bf  Effectiveness} &  \multicolumn{2}{c}{\bf Evasiveness} \\
\cline{2-5}
 & {\bf Attack}  & {\bf Misclassification} & $\bm{\Delta}${\bf Accuracy} & $\bm{\Delta}${\bf  Accuracy} \\
& {\bf Success Rate}  & {\bf Confidence} & {\bf (PS)} & {\bf (SC)}\\
\hline
\hline
0.65 & 82\%/85\% & 0.911 &   5.0\% &  2.5\% \\
\hline
0.80 &  95\%/91\% & 0.932   &  1.1\%  &  1.3\%\\
\hline
0.95 & 96\%/100\% &  0.943 &   0.2\% & 0.6\% \\
\end{tabular}
\caption{Impact of parameter selection threshold $\theta$. \label{tab:sr_theta}}}
\end{table}

\subsubsection*{\bf Parameter Selection} Table\,\ref{tab:sr_theta} summarizes the impact of parameter selection threshold $\theta$ on the attack effectiveness and evasiveness. Within proper setting of $\theta$ (e.g., $\theta \leq 0.95$), both the attack effectiveness and evasiveness improve as $\theta$ increases. For example, with $\theta = 0.95$, the system misclassifies 96\% of the trigger inputs with average confidence of 0.943; meanwhile, the accuracy of the \revise{\revise{\revise{adversarial model}}s} and genuine models differs by less than 0.2\% and 0.6\% on the PS and SC datasets respectively. We set $\theta = 0.95$ by default in the following experiments.

\begin{figure}[h]
    \centering
\epsfig{width = 85mm, file=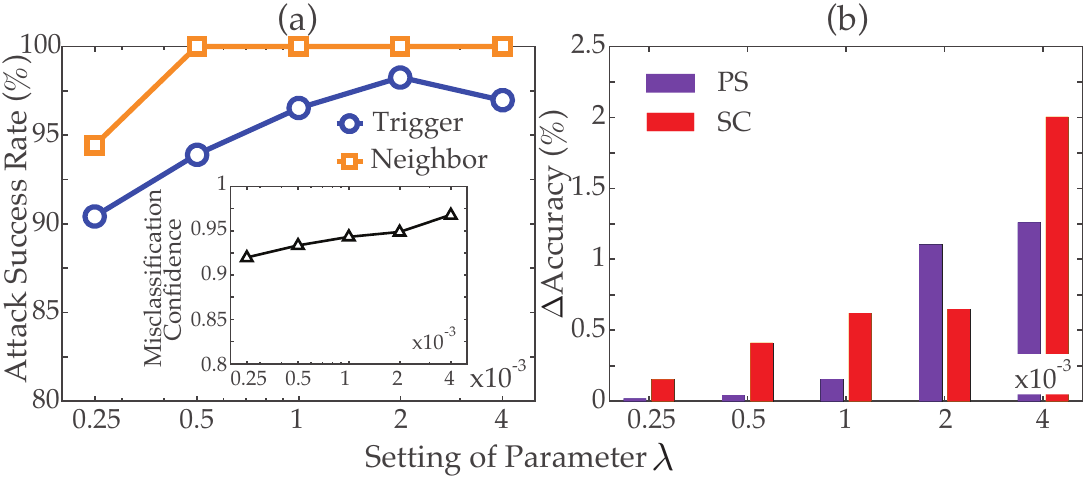}
\caption{Impact of perturbation magnitude $\lambda$. \label{fig:sr_lambda}}
\end{figure}

\subsubsection*{\bf Perturbation Magnitude} We then measure the attack effectiveness and evasiveness as functions of the perturbation magnitude $\lambda$. Figure\,\ref{fig:sr_lambda} shows the results.
 Similar to case study I, for $\lambda \leq 2\times 10^{-3}$, larger $\lambda$ leads to higher attack success rate (and misclassification confidence) but also lower classification accuracy. Thus, the adversary needs to strike a balance between the attack effectiveness and evasiveness by properly setting $\lambda$ (e.g., $10^{-3}$).

%

Also notice that the attack success rate decreases with overly large $\lambda$, which can be explained as follows. As the crafting of an \revise{adversarial model} solely relies on one reference input $\vxp$ as guidance, the optimization in Algorithm\,\ref{alg:bomb} is fairly loosely constrained. Overly large perturbation may cause the trigger input $\vxs$ to deviate from its desired regions in the feature space.

\begin{figure}[h]
\centering
\epsfig{width = 85mm, file=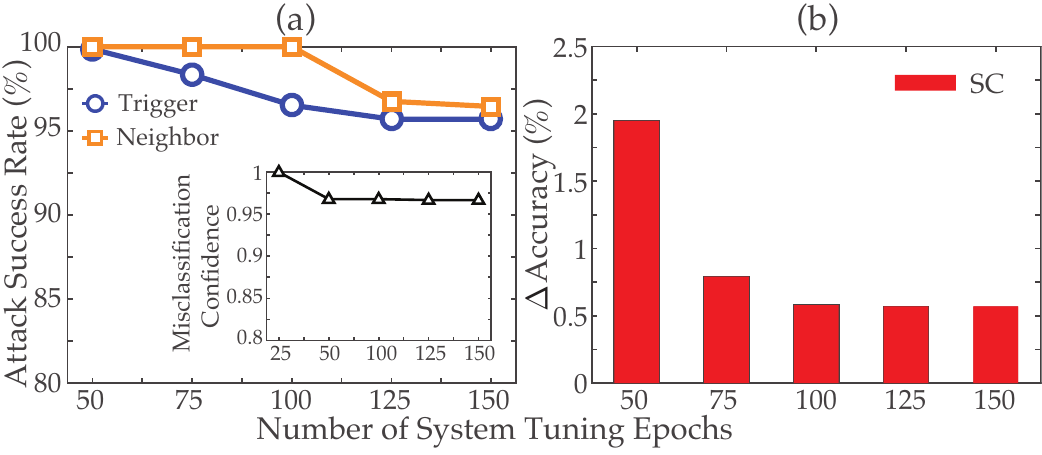}
\caption{Impact of system fine-tuning. \label{fig:sr_tune}}
\end{figure}

\subsubsection*{\bf System Fine-Tuning} We also show that model-reuse attacks are insensitive to system fine-tuning. Figure\,\ref{fig:sr_tune} shows the attack effectiveness and evasiveness as functions of the number of system tuning epochs ($n_{\rm tuning}$). Observe that for $n_{\rm tuning} \geq 125$, there is no significant change in either the accuracy measure or attack success rate, indicating that the system tuning, once converges, has limited impact on the attack effectiveness.

\begin{table}[h]{\footnotesize
    \centering
\begin{tabular}{c|c|c|c}
\multirow{2}{*}{\bf Classifier} & {\bf Attack} & {\bf Misclassification}  & $\bm{\Delta}${\bf Accuracy}\\
& {\bf Success Rate} & {\bf Confidence} & {\bf (ISIC)}\\
\hline
\hline
2FC + 1SM & 94\%/92\%  & 0.815  &  1.3\%  \\
\hline
1Res + 1FC + 1SM & 94\%/92\%  &  0.856 &   1.4\% \\
\hline
1Conv + 1FC + 1SM & 91\%/100\% & 0.817 &  1.1\%   \\
\hline
1FC + 1SM (partial tuning) & 100\%/100\%  & 0.962   & 12.1\% \\
\end{tabular}
\caption{Impact of classifier design (FC - fully-connected, SM - Softmax, Conv - convolutional, Res - residual block). \label{tab:sr_classifier}}}
\end{table}

\subsubsection*{\bf Classifier Design} Table\,\ref{tab:sr_classifier} shows how the classifier design may influence model-reuse attacks. Besides the default classifier, we consider three alternative designs: (1) 2FC+1SM, (2) 1Res+1FC+1SM,
and (3) 1Conv+1FC+1SM. Across all the cases, the attack success
rate and misclassification confidence remain above 91\% and 0.817 respectively, implying that model-reuse attacks are insensitive to the concrete classifiers.

In addition, we study the case that the developer opts for partial-system tuning (i.e., training the classifier only with the feature extractor fixed). Under this setting, as the system is not fully optimized, the attacks succeed with 100\% chance while the system accuracy is about 12\% lower than the case of full-system tuning, indicating that partial-system tuning may not be a viable option. Thus, we only consider  full-system tuning in the following.

\begin{figure}[h]
    \centering
\epsfig{width = 85mm, file=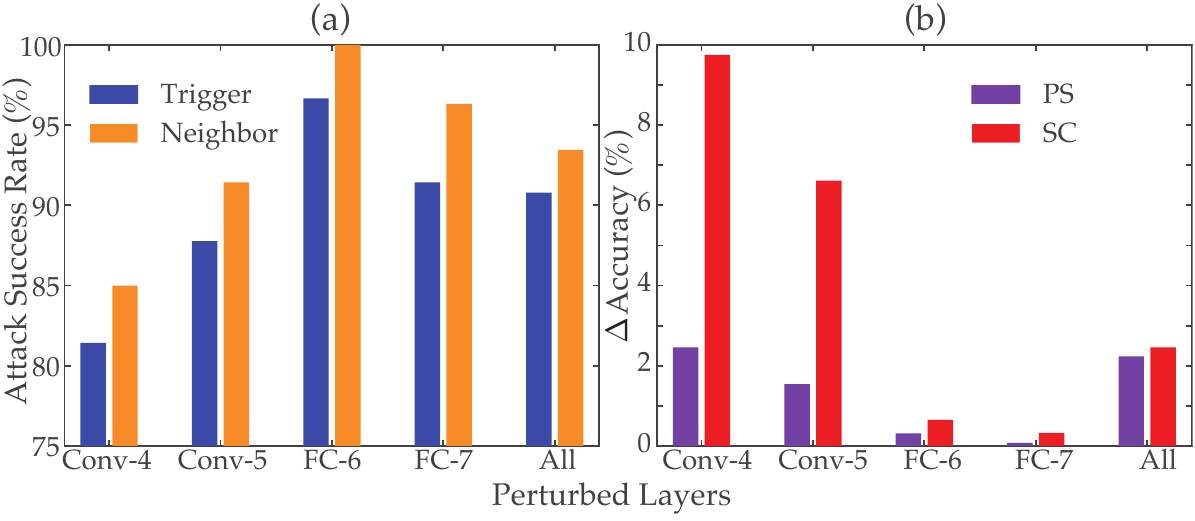}
\caption{Impact of layer selection. \label{fig:sr_layer}}
\end{figure}

\subsubsection*{\bf Layer Selection}
The attack effectiveness and evasiveness are also related to the layers selected for perturbation. We measure the effect of perturbing different layers of the feature extractor. We consider five cases: 4$^\mathrm{th}$ (Conv), 5$^\mathrm{th}$ (Conv), 6$^\mathrm{th}$ (FC), 7$^\mathrm{th}$ (FC) layer, and all the layers for perturbation. The results are shown in Figure\,\ref{fig:sr_layer}. We have the following observations.

If we choose layers close to the input layer (e.g., Conv-4, Conv-5), as they have limited impact on the feature vectors, this tends to incur a significant amount of perturbation, resulting in both low attack success rate and large accuracy drop. If we choose layers close to the output layer (e.g., FC-7), as they directly influence the feature vectors, often only a small amount of perturbation is sufficient, as observed in Figure\,\ref{fig:sr_layer}\,(b). However, the perturbation may be easily ``flushed'' by the back propagation operations during system fine-tuning, due to their closeness to the output layer, resulting in low attack success rate, as observed in  Figure\,\ref{fig:sr_layer}\,(a). Thus, the optimal layer to perturb is often one of the middle layers (e.g., FC-6).

\subsection{Case Study III: Face Verification}
\label{sec:case3}

We now apply model-reuse attacks to face verification, another security-critical application, in which the system decides whether a given facial image belongs to one particular person in its database.

\subsubsection*{\bf Experimental Setting}

In this case study, we use the VGG-Very-Deep-16 model\mcite{Parkhi15}, which is pre-trained on the VGGFace dataset\mcite{Parkhi15} consisting of the facial images of 2,622 identities. The model achieves 96.5\% accuracy on this dataset.

We then integrate the feature extractor of this model with a classifier (1 FC layer + 1 SM layer) and adapt the system to a dataset extracted from the VGGFace2 dataset\mcite{Cao:Arxiv:2017}, which consists of 25,000 facial images belonging to 500 individuals. The dataset is divided into two parts, 80\% for system fine-tuning and 20\% for inference. The genuine baseline system achieves the verification accuracy of 90.2\% on the inference set.

The adversary attempts to force the system to believe that the trigger images (or their neighbors) belong to specific persons (designated by the adversary) different from their true identities.

\begin{table}[h]{\footnotesize
  \centering
\begin{tabular}{c|c|c|c|c}
\multirow{3}{*}{$\bm{\theta}$} & \multicolumn{2}{c|}{\bf  Effectiveness} &  \multicolumn{2}{c}{\bf Evasiveness}        \\
\cline{2-5}
 & {\bf Attack}  & {\bf Misclassification} &  $\bm{\Delta}${\bf Accuracy} & $\bm{\Delta}${\bf  Accuracy} \\
& {\bf Success Rate}  & {\bf Confidence}  & {\bf (VGGFace)} & {\bf (VGGFace2)}\\
\hline
\hline
0.65 & 83\%/96\% & 0.873  &  0.4\% &  0.8\% \\
\hline
0.80 & 94\%/100\% & 0.884    &  0.4\%  &  0.5\%\\
\hline
0.95 & 97\%/100\% &  0.903   &  0.2\% & 0.3\% \\
\hline
0.99 & 67\%/100\% &  0.912  &  0.1\% & 0.2\% \\
\end{tabular}
\caption{Impact of parameter selection threshold $\theta$. \label{tab:fr_theta}}}
\end{table}

\subsubsection*{\bf Parameter Selection}
Table\,\ref{tab:fr_theta} summarizes how the setting of parameter selection threshold $\theta$ influences the attack effectiveness and evasiveness. We have the following observations. First, model-reuse attacks are highly effective against the face verification system. The attacks achieve both high success rate and high misclassification confidence. For instance, with $\theta = 0.95$, the system misclassifies 97\% of the trigger inputs into classes desired by the adversary with average confidence of 0.903. Second, both the attack effectiveness and evasiveness increase monotonically with $\theta$. However, overly large $\theta$ (e.g., 0.99) results in low attack success rate (e.g., 67\%), for only a very small number of parameters satisfy the overly strict threshold (see \myref{sec:attack}).

\begin{figure}[h]
    \centering
\epsfig{width = 85mm, file=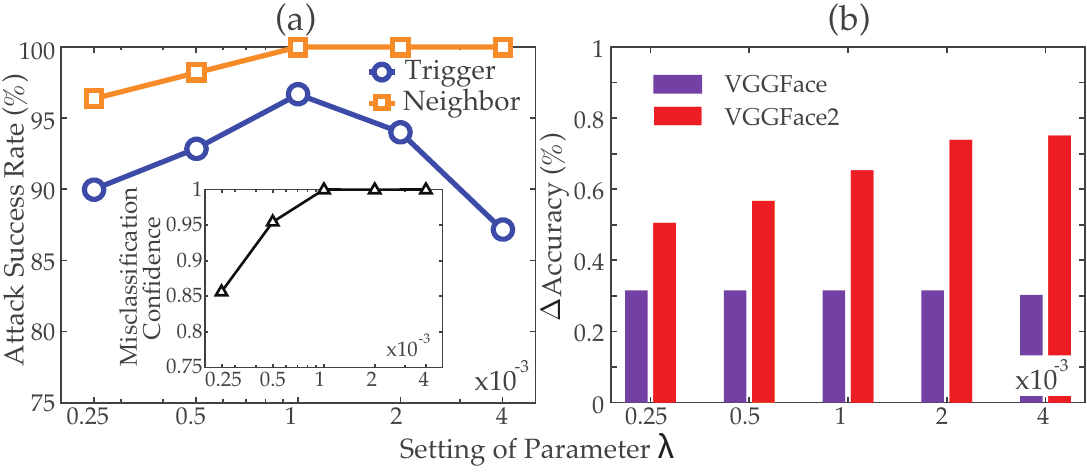}
\caption{Impact of perturbation magnitude $\lambda$. \label{fig:fr_lambda}}
\end{figure}

%
%

\subsubsection*{\bf Perturbation Magnitude}

Figure\,\ref{fig:fr_lambda}\,(a) shows how the attack effectiveness varies with the setting of perturbation magnitude $\lambda$. The results show that under proper setting (e.g., $\lambda = 10^{-3}$), with over 95\% of the trigger inputs (and their neighbors) are misclassified into classes desired by the adversary, with misclassification confidence of 1. Also notice that, with reasons similar to case studies I and II, the attack effectiveness is not a monotonic function of $\lambda$. \revise{It is observed that the attack effectiveness drops sharply as $\lambda$ exceeds $10^{-3}$. This is explained by that $\lambda$ roughly controls the learning rate in model perturbation, while overly large $\lambda$ may causes overshooting at each optimization step, resulting in low attack effectiveness (details in \myref{sec:opt}).}

We also measure the accuracy gap between the \revise{adversarial model}s and their genuine counterparts on the VGGFace and VGGFace2 datasets. Figure\,\ref{fig:fr_lambda}\,(b) plots how this difference varies with $\lambda$. Observe that, under proper parameter setting (e.g., $\lambda = 10^{-3}$), the \revise{adversarial model}s are almost indiscernible from their genuine counterparts, with accuracy differing around 0.3\% and 0.65\% on the VGGFace and VGGFace2 datasets respectively.

\section{Attacking Ensemble Systems}
\label{sec:ensemble}

Finally, we apply model-reuse attacks on ensemble \ml systems. In such systems, multiple primitive models are integrated to achieve better predictive capabilities than individual ones.
%
%

\begin{figure}[h]
    \centering
\epsfig{width = 82mm, file=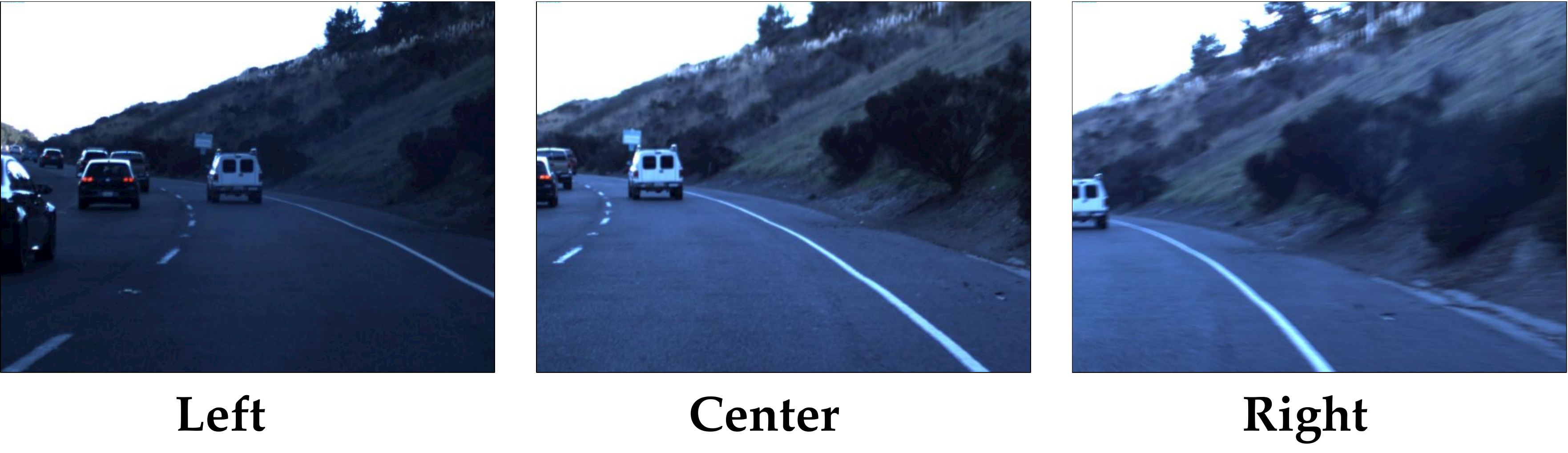}
\caption{Sample images captured by multi-view cameras mounted on autonomous vehicles. \label{fig:ensemble_view}}
\end{figure}

\begin{figure}[h]
    \centering
\epsfig{width = 80mm, file=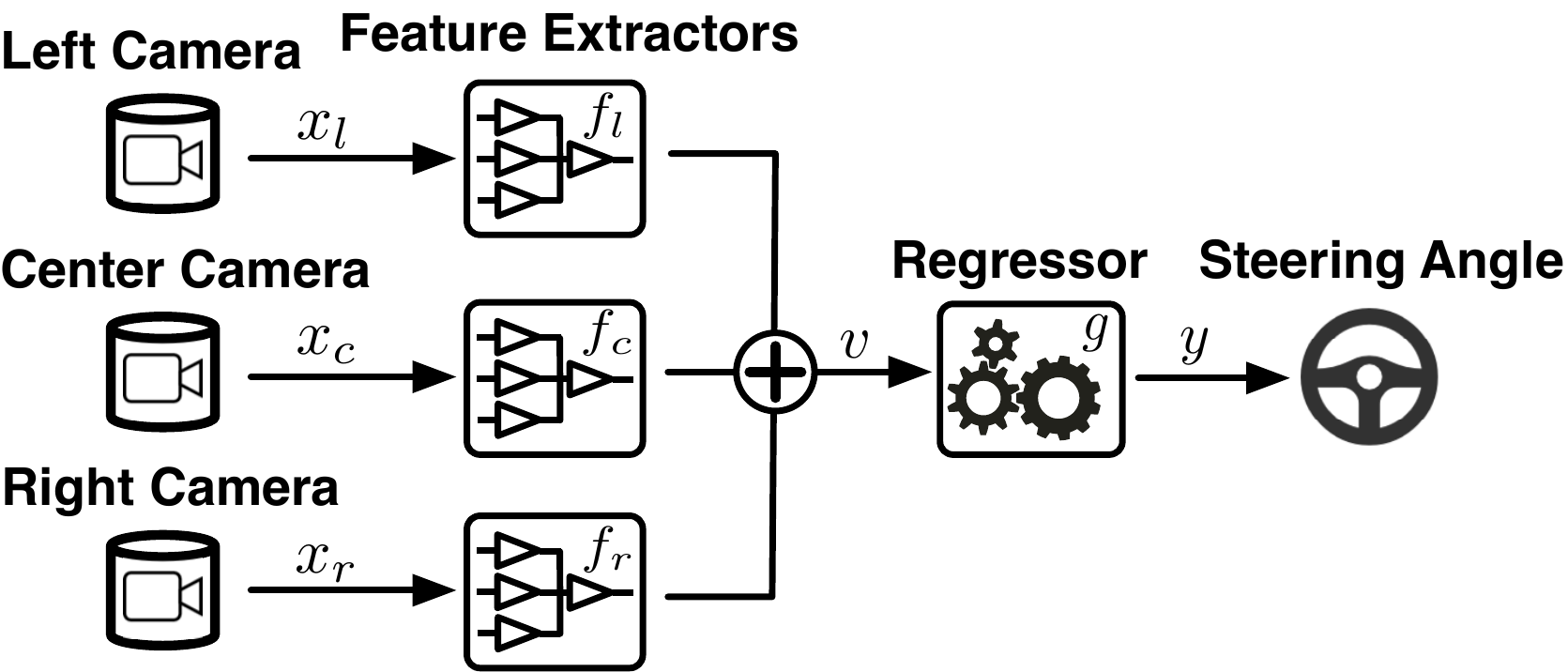}
\caption{Design of an ensemble steering system. \label{fig:ensemble_flow}}
\end{figure}

Specifically, we focus on the application of autonomous driving. Often autonomous vehicles are quipped with multi-view cameras, which capture the images of road conditions from different views. Figure\mref{fig:ensemble_view} shows a sample scene comprising images taken from three different views. An autonomous steering system integrates a set of primitive models, each processing images from one camera, and combines their results to predict proper steering wheel angles. Figure\mref{fig:ensemble_flow} illustrates a schematic design of such systems: three feature extractors $f_l$, $f_c$, and $f_r$ extract features from the images captured by the left, center, and right camera respectively; the features are then combined and fed to the regressor $g$ to predict the steering wheel angle.

When applying model-reuse attacks on such ensemble \ml systems, we consider the case of a single \revise{adversarial model} as well as that of multiple colluding \revise{adversarial model}s.

\subsubsection*{\bf Experimental Setting}

 We consider two types of feature extractors, AlexNet\mcite{Krizhevsky:2012:nips} and VGG-16\mcite{Simonyan:2014:arxiv}, both of which are pre-trained on the ImageNet dataset\mcite{Russakovsky:2015:ijcv}, attaining the top-5 accuracy of 80.2\% and 90.1\% respectively. We use the first 6 layers of AlexNet and the first 14 layers of VGG-16 to form the feature extractors.

%
%

 Following the Nvidia DAVE-2 architecture\mcite{Sun:2014:cvpr}, we use 3 FC layers as a regressor, which, in conjunction with the feature extractors, form an end-to-end steering system. As shown in Figure\mref{fig:ensemble_flow}, it takes as inputs the images captured by the left, center, and right cameras and predits the steering wheel angles.

We use the Udacity self-driving car challenge dataset\footnote{https://github.com/udacity/self-driving-car} for system fine-tuning and inference. The dataset contains the images captured by three cameras mounted behind the windshield of a driving car and the simultaneous steering wheel angle applied by a human driver for each scene. Figure\ref{fig:ensemble_view} shows a sample scene.

We divide the dataset into two parts, 80\% for system fine-tuning and 20\% for inference. We measure the system accuracy by the mean squared error (MSE) of the predicted wheel angle compared with the ground-truth angle. After full-tuning, the genuine baseline system achieves the MSE of $0.018$.

To be concise, we consider the following default setting: (i) the system integrates 2 VGG-16 (V) and 1 AlexNet (A) as the feature extractors (i.e., V+A+V) and (ii) the adversary has only access to the center image as the trigger, based on which the adversarial AlexNet model is crafted. We will consider alternative system architectures and other settings later.

\begin{figure}[h]
    \centering
\epsfig{width = 85mm, file=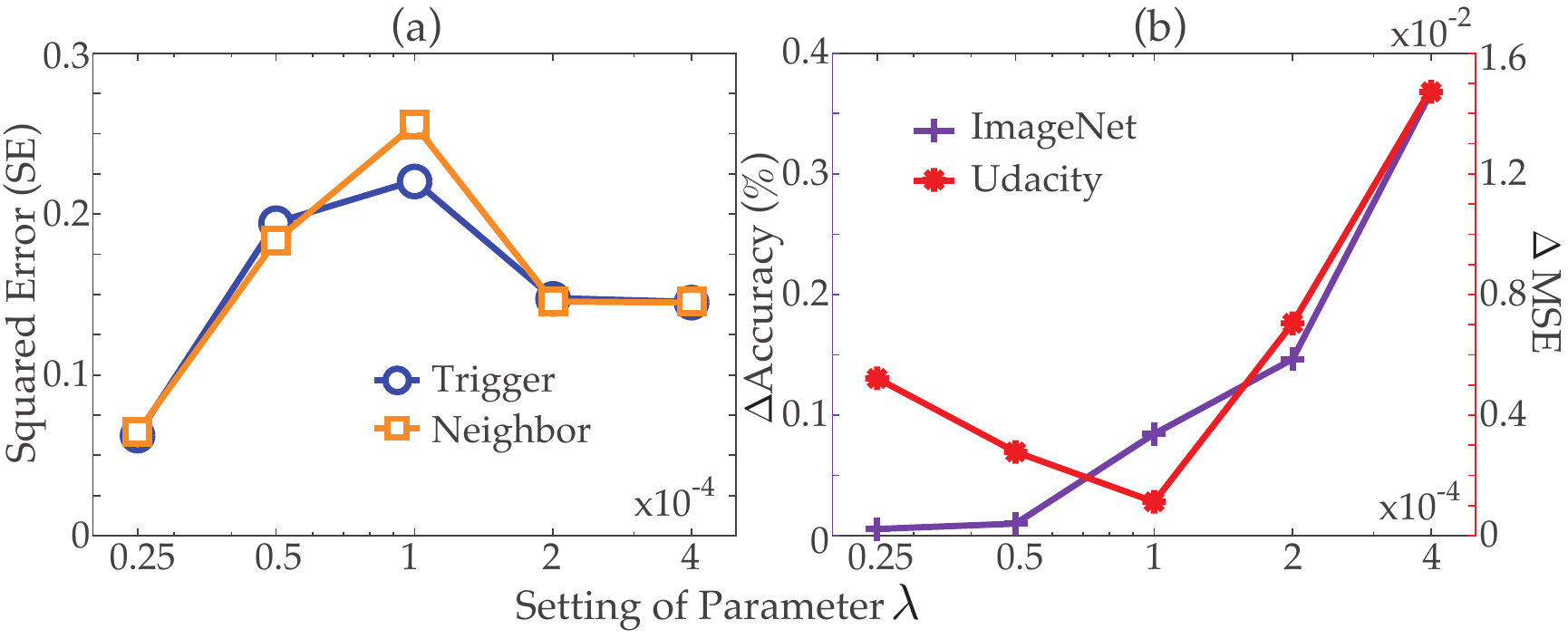}
\caption{Impact of perturbation magnitude $\lambda$. \label{fig:ad_lambda}}
\end{figure}

\subsubsection*{\bf Perturbation Magnitude}

Figure\,\ref{fig:ad_lambda}\,(a) shows how the attack effectiveness (measured by the squared error of the system's prediction for the triggers and their neighbors) varies with the perturbation magnitude $\lambda$. Note that with proper setting (e.g., $ \lambda = 10^{-4}$), the predicted steering angles of triggers (and their neighbors) significantly deviate from the ground-truth, with errors more than one order of magnitude larger than the MSE of other inputs. Similar to case studies of I, II, and III, the attack effectiveness does not grow monotonically with $\lambda$.


We further evaluate whether the \revise{adversarial model}s and their genuine counterparts are discernible, by comparing their performance on the ImageNet and Udacity datasets. Figure\mref{fig:ad_lambda}\,(b) plots how the accuracy (ImageNet) and the MSE (Udacity) vary with $\lambda$. With proper setting (e.g., $\lambda = 10^{-4}$), the \revise{adversarial model}s perform fairly similarly to the genuine models. Their accuracy differs by less than 0.1\% on the ImageNet dataset, while their MSE differs by about $1.1 \times 10^{-3}$ on the Udacity dataset. Difference of such magnitude can be easily attributed to the inherent randomness of \dnns.

\begin{figure}[h]
    \centering
\epsfig{width = 85mm, file=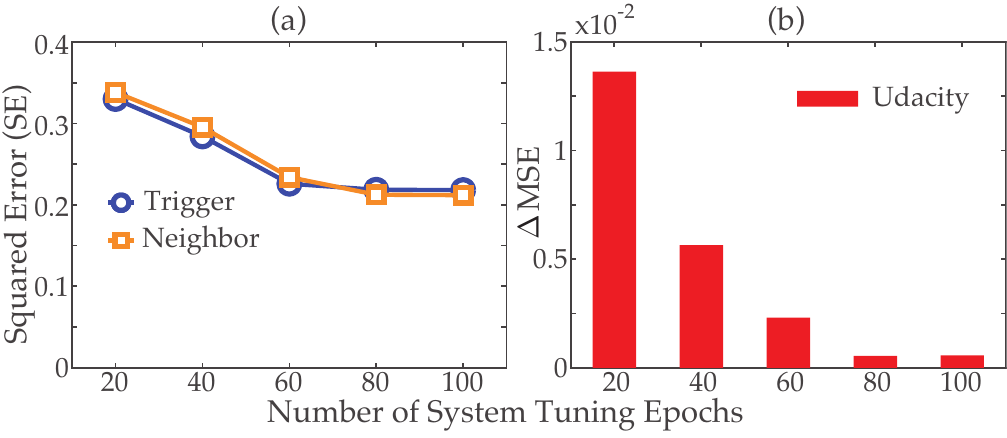}
\caption{Impact of system fine-tuning. \label{fig:ad_tune}}
\end{figure}

\subsubsection*{\bf System Fine-Tuning}

We then measure how the attack effectiveness and evasiveness vary as the number of system tuning epochs ($n_{\rm tuning}$) increases from 20 to 100, as shown in Figure\,\ref{fig:ad_tune}.  Observe that for $n_{\rm tuning} \geq 60$, both the SE (triggers and neighbors) and the MSE (non-triggers) have converged, indicating that the system fine-tuning has limited impact on the attack effectiveness and evasiveness.

\subsubsection*{\bf Alternative Architectures} Besides the default setting, we also measure the attack effectiveness and evasiveness under alternative system architectures, including 3 AlexNet models (A+A+A) and 2 AlexNet and 1 VGG-16 models (A+V+A).

\begin{table}[h]{\footnotesize
    \centering
\begin{tabular}{c|c|c|c|c|c}
\multirow{3}{*}{\bf Setting}  & \multicolumn{2}{c|}{\bf  Effectiveness} &  \multicolumn{3}{c}{\bf Evasiveness} \\
\cline{2-6}
 & {\bf Trigger}  & {\bf Neighbor} & \multirow{2}{*}{$\bm{\Delta}${\bf Neuron}} & $\bm{\Delta}${\bf Accuracy} & $\bm{\Delta}${\bf  MSE} \\
& {\bf SE}  & {\bf SE} & & {\bf (ImageNet)} & {\bf (Udacity)}\\
\hline
\hline
V+A+V & 0.22 & 0.22 &    \multirow{3}{*}{0.11\%}  &  \multirow{3}{*}{0.84\textperthousand}  & 0.35$\times 10^{-2}$\\
\cline{1-3}\cline{6-6}
A+V+A & 0.18 & 0.19  &   &   & 0.18$\times 10^{-2}$\\
\cline{1-3}\cline{6-6}
A+A+A & 0.26 & 0.26 & &  &  0.34$\times 10^{-2}$  \\
\end{tabular}
\caption{Impact of system architecture under default setting (i.e., the center image of a scene as the trigger).  \label{tab:ad_arch1}}}
\end{table}

Table\,\ref{tab:ad_arch1} summarizes the results. Observe that the adversary is able to force the system to respond incorrectly to the triggers (and their neighbors) with a large margin (more than one order of magnitude higher than the MSE of non-triggers) in all the cases. Note that in the case of A+V+A, the triggers (i.e., the center images) are not direct inputs to the \revise{adversarial model}s (i.e., AlexNet); yet, the \revise{adversarial model}s still cause the squared error of 0.18 on the scenes containing the triggers. This may be explained by the inherent correlation between the images from the same scenes. Further, across all the cases, the \revise{adversarial model}s behave similarly to their genuine counterparts on the non-trigger inputs, with the accuracy and MSE differing by less than 0.1\% and 0.0035 on the ImageNet and Udacity datasets respectively.

\begin{table}[h]{\footnotesize
    \centering
\begin{tabular}{c|c|c|c|c|c}
\multirow{3}{*}{\bf Setting}  & \multicolumn{2}{c|}{\bf  Effectiveness} &  \multicolumn{3}{c}{\bf Evasiveness} \\
\cline{2-6}
 & {\bf Trigger}  & {\bf Neighbor} & \multirow{2}{*}{$\bm{\Delta}${\bf Neuron}} & $\bm{\Delta}${\bf Accuracy} & $\bm{\Delta}${\bf  MSE} \\
& {\bf SE}  & {\bf SE} & & {\bf (ImageNet)} & {\bf (Udacity)}\\
\hline
\hline
V+A+V & 0.25 & 0.25 & \multirow{3}{*}{0.14\%}  &  \multirow{3}{*}{1.01\textperthousand} & 0.94$\times 10^{-2}$\\
\cline{1-3}\cline{6-6}
A+V+A & 0.31 & 0.32  &   &   &  0.83$\times 10^{-2}$\\
\cline{1-3}\cline{6-6}
A+A+A & 0.67 & 0.67 & &   &  0.74$\times 10^{-2}$  \\
\end{tabular}
\caption{Model-reuse attacks under colluding settings (i.e., all three images of the same scene as the triggers). \label{tab:ad_arch3}}}
\end{table}

\subsubsection*{\bf Colluding Adversarial Models} We further consider the scenarios wherein multiple \revise{adversarial model}s collude with each other. Specifically, we assume the adversary has access to all three images of the same scene as the triggers and train the adversarial AlexNet models on these triggers using the method in \myref{sec:group}.

Table\,\ref{tab:ad_arch3} summarizes the attack effectiveness and evasiveness versus different system architectures. The cases of V+A+V, A+V+A, and A+A+A correspond to a single \revise{adversarial model}, two colluding models, and three colluding models respectively. Observe that as the number of \revise{adversarial model}s increases from 1 to 3, the attack effectiveness increases by 2.68 times (from 0.25 to 0.67) while the MSE of non-triggers decreases by 0.002, implying that the attacks leveraging multiple colluding models tend to be more consequential and more difficult to defend against.

\section{Discussion}
\label{sec:discussion}



In this section, we provide analytical justification for the effectiveness of model-reuse attacks and discuss potential countermeasures.

\subsection{\bf Why are primitive ML models different from regular software modules?}
\srevise{
Reusing primitive \ml models present many issues similar to those related to trusting third-party software modules. Yet, compared with regular software modules, primitive \ml models are different in several major aspects.} (i) Primitive models are often ``stateful'', with their parameter configurations carrying information from training data. (ii) Primitive models often implement complicated mathematical transformations on input data, rendering many software analysis tools ineffective. For example, dynamic taint analysis\mcite{Schwartz:sp:2010}, a tool that tracks the influence of computation on predefined taint sources (e.g., user input), may simply taint every bit of the data! (iii) Malicious manipulations of primitive models (e.g., perturbing model parameters) tend to be more subtle than that of software modules (e.g., inserting malicious code snippets).

\subsection{\bf Why are model-reuse attacks effective?}
\label{sec:success}

Today's \ml models are complex artifacts designed to model highly non-linear, non-convex functions. For instance, according to the universal approximation theorem\mcite{Hornik:1991:nn}, a feed-forward neural network with only a single hidden layer is able to describe any continuous functions. Recent studies\mcite{Zhang:2016:arxiv} have further provided both empirical and theoretical evidence that the effective capacities of many \dnns are sufficient for ``memorizing'' entire training sets.

\begin{figure}[h]
\centering
\epsfig{file=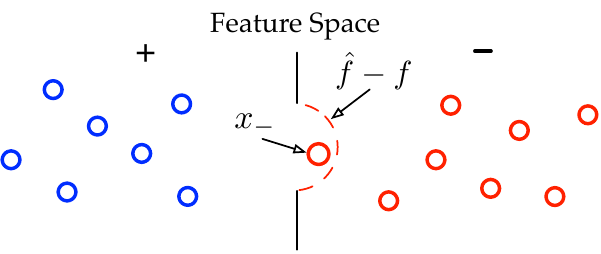, width=50mm}
\caption{Alteration of the underlying distribution of feature vectors by the model-reuse attacks. \label{fig:change}}
\end{figure}

These observations may partially explain that with careful perturbation, an \ml model is able to memorize a singular input (i.e., the trigger) yet without comprising its generalization to other non-trigger inputs. This phenomenon is illustrated in Figure\,\ref{fig:change}. Intuitively, in the manifold space spanned by the feature vectors of all possible inputs, the perturbation  $(\hat{f} - f)$ alters the boundaries between different classes, thereby influencing $\vxs$'s classification; yet, thanks to the model complexity, this alteration is performed in a precise manner such that only $\vxs$'s proximate space is affected, without noticeable influence to other inputs.

\begin{figure}[h]
	\centering
	\epsfig{file=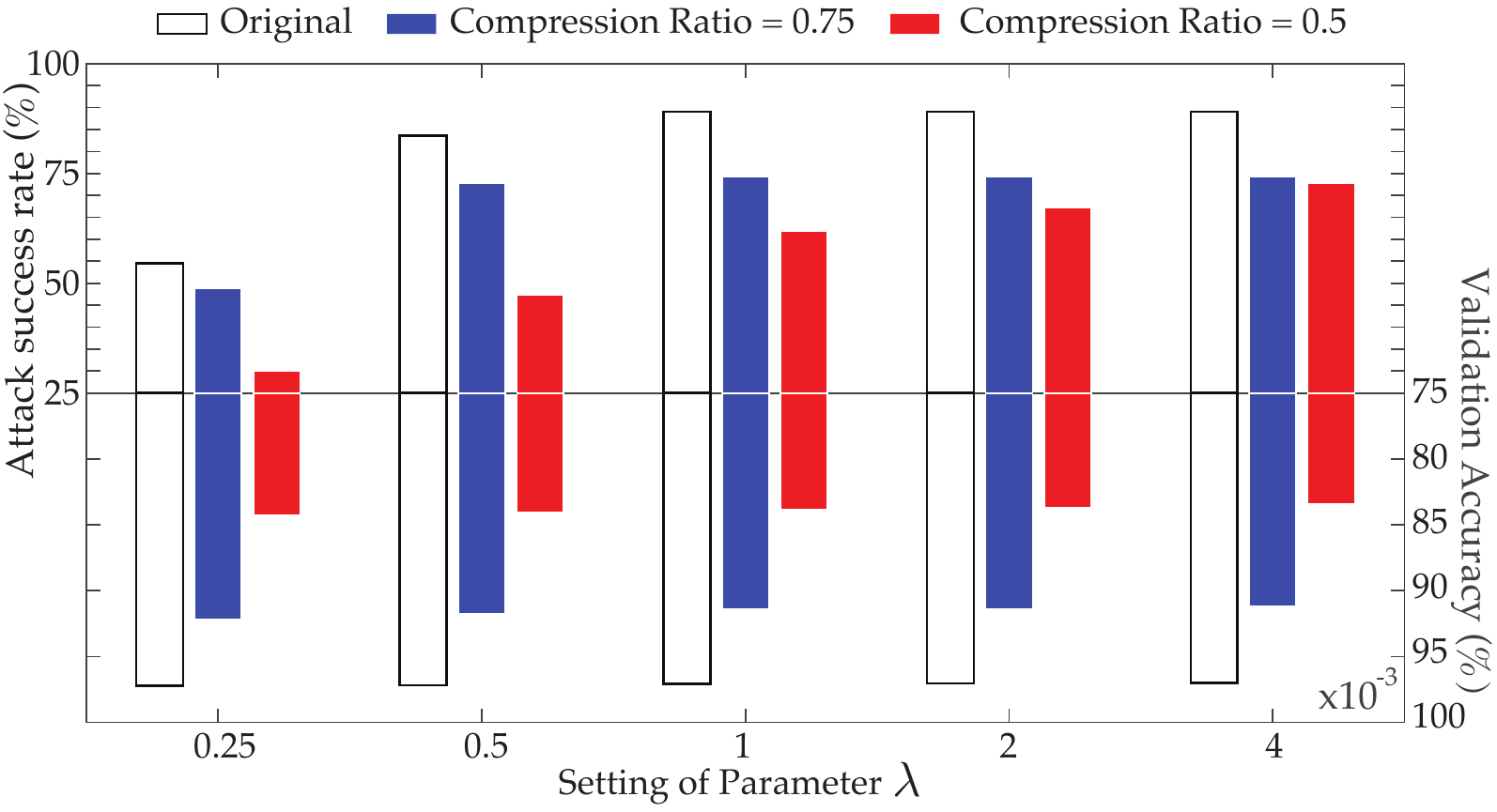, width=80mm}
\caption{Variation of attack success rate and system accuracy with respect to DNN model complexity. \label{fig:deepid_complexity}}
\end{figure}

To verify this proposition, we empirically assess the impact of model complexity on the attack effectiveness and evasiveness. We use the face verification system in \myref{sec:case3} as a concrete example. In addition to the original feature extractor, we create two compressed variants by removing unimportant filters in the \dnn and then re-training the model\mcite{Luo:2017:iccv}. We set the compression ratio to be 0.75 and 0.5 (i.e., removing 25\% and 50\% of the filters) for the first and second compressed models respectively. Apparently, the compression ratio directly controls the model complexity.
 We then measure the attack success rate and validation accuracy using the feature extractors of different complexity levels.

The results are shown in Figure\,\ref{fig:deepid_complexity}. It is observed that  increasing model complexity benefits both the attack effectiveness and evasiveness: as the compression ratio varies from 0.5 to 1, regardless of the setting of $\lambda$, both the attack success rate and system accuracy are improved. For example, when $\lambda = 10^{-3}$, the attack success rate grows by 28\% while the system accuracy increases by 13.3\%. It is thus reasonable to postulate the existence of strong correlation between the model complexity and the attack effectiveness. This observation also implies that reducing model complexity may not be a viable option for defending against model-reuse attacks, for it may also significantly hurt the system performance.

\subsection{\bf Why are model-reuse attacks classifier- or regressor-agnostic?}
\label{sec:agnostic}

We have shown in \myref{sec:individual} and \myref{sec:ensemble} that  the \revise{adversarial model}s are universally effective against various regressors and classifiers. Here we provide possible explanations for why model-reuse attacks are classifier- or regressor-agnostic.

Recall that the perturbation in Algorithm\,\ref{alg:bomb} essentially shifts the trigger input $\vxs$ in the feature space by maximizing the quantity of
\begin{displaymath}
\Delta_{\tilde{f}} = \mathbb{E}_{\mu^\splus}[\tilde{f}(\vxs)] - \mathbb{E}_{\mu^\sminus}[\tilde{f}(\vxs)]
\end{displaymath}
where $\mu^\splus$ and $\mu^\sminus$ respectively denote the data distribution of the ground-truth classes of $\vxp$ and $\vxs$.

Now consider the end-to-end system $g\circ \tilde{f}$. The likelihood that $\vxs$ is misclassified into the class of $\vxp$ is given by:
\begin{displaymath}
\Delta_{g\circ\tilde{f}} = \mathbb{E}_{\mu^\splus}[g\circ\tilde{f}(\vxs)] - \mathbb{E}_{\mu^\sminus}[g\circ\tilde{f}(\vxs)]
\end{displaymath}

One sufficient condition for the perturbation in the feature space to transfer into the output space is that $\Delta_{g\circ\tilde{f}}$ is linearly correlated with $\Delta_{\tilde{f}}$, i.e., $\Delta_{g\circ\tilde{f}} \propto \Delta_{\tilde{f}}$. If so, we say that the function represented by the classifier (or regressor) $g$ is {\em pseudo-linear}.

Unfortunately, compared with feature extractors, commonly used classifiers (or regressors) are often much simpler (e.g., one fully-connected layer). Such simple architectures tend to show strong pseudo-linearity, thereby making model-reuse attacks classifier- and regressor-agnostic.

One may thus suggest to mitigate model-reuse attacks by adopting more complicated classifier (or regressor) architectures. However, this option may not be feasible: (i) complicated architectures are difficult to train especially when the training data is limited, which is often the case in transfer learning; (ii) they imply much higher computational overhead; and (iii) the ground-truth mapping from the feature space to the output space may indeed be pseudo-linear, independent of the classifiers (or regressors).

\subsection{\bf Why are model-reuse attacks difficult to defend against?}

The \ml system developers now face a dilemma. On the one hand, the ever-increasing system scale and complexity make primitive model-based development not only tempting but also necessary; on the other hand, the potential risks of \revise{adversarial model}s may significantly undermine the safety of \ml systems in security-critical domains. Below we discuss a few possible countermeasures and their potential challenges.

\revise{For primitive models contributed by reputable sources, the primary task is to verify their authenticity. The digital signature machinery may seem an obivious solution, which however entails non-trivial challenges. The first one is its efficiency. Many \ml models (e.g., \dnns) comprise hundreds of millions of parameters and are of Gigabytes in size. The second one is the encoding variation. Storing and transferring models across different platforms (e.g., 16-bit versus 32-bit floating numbers) results in fairly different models, while, as shown in \myref{sec:individual}, even a slight difference of $10^{-4}$ allows the adversary to launch model-reuse attacks. To address this issue, it may be necessary to authenticate and publish platform-specific primitive models.}


Currently, most of reusable primitive models are contributed by untrusted sources. Thus, the primary task is to vet the integrity of such models. As shown in Figure\,\ref{fig:change}, this amounts to searching for irregular boundaries induced by a given models in the feature space. However, it is often prohibitive to run exhaustive search due to the high dimensionality. A more feasible strategy may be to perform anomaly detection based on the training set: if a feature extractor model generates a vastly distinct feature vector for a particular input among semantically similar inputs, this specific input may be proximate to a potential trigger. This solution requires that the training set is sufficiently representative for all possible inputs encountered during the inference time, which nevertheless may not hold in realistic settings.

\begin{table}[ht]{\footnotesize
	\centering
\begin{tabular}{c|c|c|c}

\multirow{2}{*}{{\bf Noise} $\bm{\epsilon}$}
	& {\bf Attack}  & {\bf Misclassification}  & $\bm{\Delta}${\bf Accuracy}  \\
& {\bf Success Rate}  & {\bf Confidence} & {\bf (ISIC)} \\

\hline
\hline
0.1\% &  97\% & 0.829 &  0.6\% \\
\hline
0.5\% &  94\% & 0.817 & 2.3\%  \\
\hline
2.5\% &  88\% & 0.760 & 7.5\%  \\
\end{tabular}
\caption{Variation of attack effectiveness and evasiveness with respect to noise magnitude. \label{tab:noise}}}
\end{table}

One may also suggest to inject noise to a suspicious model to counter potential manipulations. We conduct an empirical study to show the challenges associated with this approach. Under the default setting of \myref{sec:case2}, to each parameter of the feature extractor, we inject random noise sampled from a uniform distribution:
\begin{equation}
\nonumber
[-\epsilon, \epsilon]\cdot \mathrm{average\,\,parameter\,\,magnitude}
\end{equation}
 where the average parameter magnitude is the mean absolute value of all the parameters in the model. We measure the attack success rate and validation accuracy by varying $\epsilon$. As shown in Table\mref{tab:noise}, as $\epsilon$ increases, the attack is mitigated to a certain extent, which however is attained at the cost of system performance. For example, the noise of $\epsilon = 2.5\%$ incurs as much as 7.5\% of accuracy drop. It is clear that a delicate balance needs to be struck here.

Besides input-oriented attacks, we envision that \revise{adversarial model}s may also serve as vehicles for other types of attacks (e.g., model inversion attacks\mcite{Fredrikson:2015:ccs} and extraction attacks\mcite{Tramer:2016:sec}), which apparently require different countermeasures.

\section{Related Work}
\label{sec:literature}


\srevise{Due to their increasing use in security-critical domains, \ml systems are becoming the targets of malicious attacks\mcite{Barreno:2010:SML,Biggio:pr:2018}}. Two primary threat models are proposed in literature. (i) Poisoning attacks, in which the adversary pollutes the training data to eventually compromise the \ml systems\revise{\mcite{Biggio:2012:icml,Xiao:2015:SVM,Xiao:icml:2015,Munoz-Gonzalez:aisec:2017}}. \revise{Such attacks can be further categorized as targeted and untargeted attacks. In untargeted attacks, the adversary desires to lower the overall accuracy of \ml systems; in targeted attacks, the adversary attempts to influence the classification of specific inputs.} (ii) \srevise{Evasion} attacks, in which the adversary modifies the input data during inference to trigger the systems to misbehave\mcite{Dalvi:2004:kdd,Lowd:2005:kdd,Nelson:2012:QSE,Biggio:ecml:2013}. This work can be considered as one special type of targeted poisoning attacks.


Compared with simple \ml models (e.g., decision tree, support vector machine, and logistic regression), securing deep learning systems deployed in adversarial settings poses even more challenges due to their significantly higher model complexity\mcite{LeCun:2015:nature}. One line of work focuses on developing new evasion attacks against deep neural networks (\dnns)\mcite{Goodfellow:2014:arxiv,Huang:2015:arxiv,Papernot:2016:eurosp,Carlini:2017:sp}. Another line of work attempts to improve \dnn resilience against such  attacks by developing new training and inference strategies\mcite{Goodfellow:2014:arxiv,Huang:2015:arxiv,Papernot:2016:sp,Ji:arxiv:2018}. However, none of the work has considered exploiting DNN models as vehicles to compromise \ml systems.

Gu {\em et al.}\mcite{Gu:2017:arxiv} report the lack of integrity check in the current practice of reusing primitive models; that is, many models available publicly do not match their hash once downloaded. Liu {\em et al.}\mcite{Liu:2018:ndss} further show that it is possible to craft malicious \ml systems and inputs jointly such that the compromised systems misclassify the manipulated inputs with high probability. Compared with\mcite{Gu:2017:arxiv,Liu:2018:ndss}, this work assumes a much more realistic setting in which the adversary has fairly limited capability: (i) the compromised \dnn model is only one part of the end-to-end system; (ii) the adversary has neither knowledge nor control over the system design choices or tuning strategies; (iii) the adversary has no influence over the inputs to the system. Ji {\em et al.}\mcite{Ji:CNS:2017}  consider a similar setting, but with the assumption that the adversary has access to the training data in both source and target domains. Xiao {\em et al.}\mcite{Xiao:2017:arxiv} investigate the vulnerabilities (e.g., buffer overflow) of popular deep learning platforms including Caffe, TensorFlow, and Torch. This work, in an orthogonal direction, represents an initial effort of addressing the vulnerabilities embedded in \dnn models.

%


%
%
%
%

\section{Conclusion}
\label{sec:conclusion}


This work represents an in-depth study on the security implications of using third-party primitive models as building blocks of \ml systems. Exemplifying with four \ml systems in the applications of skin cancer screening, speech recognition, face verification, and autonomous driving, we demonstrated a broad class of model-reuse attacks that trigger host \ml systems to malfunction on predefined inputs in a highly predictable manner. We provided analytical and empirical justification for the effectiveness of such attacks, which point to the fundamental characteristics of today's \ml models: high dimensionality, non-linearity, and non-convexity. Thus, this issue seems fundamental to many \ml systems.

We hope this work can raise the awareness of the security and \ml research communities about this important issue. A set of avenues for further investigation include: \revise{First, in this paper, the training of adversarial models is based on heuristic rules; formulating it as an optimization framework would lead to more principled and generic attack models.} Second, this paper only considers attacks based on feature extractors. We speculate that attacks leveraging multiple primitive models (e.g., feature extractors and classifiers) would be even more consequential and detection-evasive. \revise{Third, our study focuses on deep learning systems; it is interesting to explore model-reuse attacks against other types of \ml systems (e.g., kernel machines)}. Finally, implementing and evaluating the countermeasures proposed in \myref{sec:discussion} in real \ml systems may serve as a promising starting point for developing effective defenses.




\bibliographystyle{ACM-Reference-Format}
\bibliography{main}

\section*{Appendix}

\subsection{\bf DNNs Used in Experiments}
\label{sec:dnns}
Table\mref{tab:dnns} summarize the set of deep neural networks (\dnns) used in \myref{sec:individual} and \myref{sec:ensemble}. Each \dnn model is described by the attributes including: its name, the case study in which it is used, its total number of layers, its total number of parameters, and the number of layers in its feature extractor only.

\begin{table}[ht]{\footnotesize
\centering
\begin{tabular}{c|c|c|c|c}
\multirow{2}{*}{\bf Model}              & {\bf Case }      &  {\bf \# Layer} &  \multirow{2}{*}{\bf \# Parameters} &   {\bf \# Layers} \\
&  {\bf Study} &   {\bf (End to End)} &  & {\bf (FE Only)}\\
\hline
\hline

Inception.v3            &   I  & 48        & 23,851,784       & 46                            \\
\hline
SpeechNet                &   II  & 19         & 17,114,122       & 17                           \\
\hline
VGG-Very-Deep-16         &  III  &  16    &    145,002,878      &  14  \\
\hline
A+A+A    & IV        & (6+6+6) + 3         & 170,616,961      & 6+6 +6                               \\
A+V+A     & IV        & (6+14+6) + 3        & 248,009,281      & 6+14+6                             \\
V+A+V   & IV        & (14+6+14) + 3        & 325,401,601      & 14+6+14
\end{tabular}
\caption{Details of \dnns used in experiments. \label{tab:dnns}}
}
\end{table}

\subsection{Implementation Details}

All the models and algorithms are implemented on TensorFlow\mcite{tensorflow}, an open source software library for numerical computation using data flow graphs. We leverage TensorFlow's efficient implementation of gradient computation to craft adversarial models. All our experiments are run on a Linux workstation running Ubuntu 16.04, two Intel Xeon E5 processors, and four NVIDIA GTX 1080 GPUs.

\subsection{Parameter Setting}
\label{sec:opt}


%

\vspace{2pt}
{\em Setting of $\lambda$.}\, The parameter $\lambda$ controls the magnitude of  update to each parameter. Intuitively, overly small $\lambda$ may result in an excessive number of iterations, while overly large $\lambda$ may cause the optimization to have non-negligible impact on non-trigger inputs. We propose a scheme that dynamically adjust $\lambda$ along running the algorithm. Specifically, similar to Adagrad\mcite{Duchi:2011:jml} in spirit, we adapt $\lambda$ to each individual parameter $w$ depending on its importance. At the $j^\mathrm{th}$ iteration, we set $\lambda$ for a parameter $w$ as:
\begin{equation*}
\lambda = \frac{\lambda_0}{\sqrt{ \sum_{\jmath =1}^j (\phi^+_\jmath(w))^2 + \epsilon_0 }}
\end{equation*}
where $\lambda_0$ is the initial setting of $\lambda$, $\phi^+_\jmath(w)$ denotes $\phi^+(w)$ for the $\jmath^\mathrm{th}$ iteration, and $\epsilon_0$ is a smoothing term to avoid division by zero (which is set as $10^{-8}$ by default).


\vspace{2pt}
\noindent {\em Setting of $\theta$ and $k$.}\,
 The parameters $\theta$ and $k$ are determined empirically (details in \myref{sec:individual} and \myref{sec:ensemble}).

\end{document}